\documentclass[aps,prl,twocolumn,superscriptaddress,showpacs]{revtex4}
\usepackage[colorlinks=true,urlcolor=blue,citecolor=blue,linkcolor=blue]{hyperref}
\usepackage{graphicx}
\usepackage{mathrsfs}
\usepackage{bm}
\usepackage{verbatim}
\usepackage{hyperref}
\begin{document}
\title{Chiral Interface States and Related Quantized Transport in Disordered Chern Insulators}

\author{Zhi-Qiang Zhang}
\affiliation{School of Physical Science and Technology, Soochow University, Suzhou 215006, China}

\author{Chui-Zhen Chen}
\affiliation{Institute for Advanced Study, Soochow University, Suzhou 215006, China}
\affiliation{School of Physical Science and Technology, Soochow University, Suzhou 215006, China}

\author{Yijia Wu}
\affiliation{International Center for Quantum Materials, School of Physics, Peking University, Beijing 100871, China}

\author{Hua Jiang} \email{jianghuaphy@suda.edu.cn}
\affiliation{School of Physical Science and Technology, Soochow University, Suzhou 215006, China}
\affiliation{Institute for Advanced Study, Soochow University, Suzhou 215006, China}

\author{Junwei Liu}
\affiliation{Department of Physics, Hong Kong University of Science and Technology, Kowloon, Hong Kong}

\author{Qing-feng Sun}
\affiliation{International Center for Quantum Materials, School of Physics, Peking University, Beijing 100871, China}
\affiliation{Beijing Academy of Quantum Information Sciences, Beijing 100193, China}
\affiliation{CAS Center for Excellence in Topological Quantum Computation, University of Chinese Academy of Sciences, Beijing 100190, China}

\author{X. C. Xie}  
\affiliation{International Center for Quantum Materials, School of Physics, Peking University, Beijing 100871, China}
\affiliation{Beijing Academy of Quantum Information Sciences, Beijing 100193, China}
\affiliation{CAS Center for Excellence in Topological Quantum Computation, University of Chinese Academy of Sciences, Beijing 100190, China}

\date{\today}
\begin{abstract}
In this Letter, we study an Anderson-localization-induced quantized transport in disordered Chern insulators (CIs). By investigating the disordered CIs with a step potential, we find that the chiral interface states emerge along the interfaces of the step potential, and the energy range for such quantized transport can be manipulated through the potential strength. Furthermore, numerical simulations on the case with a multi-step potential demonstrate that such chiral state can be spatially shifted by varying the Fermi energy, and the energy window for quantized transport is greatly enlarged. Experimentally, such chiral interface states can be realized by imposing transverse electric field, in which the energy window for quantized transport is much broader than the intrinsic band gap of the corresponding CI. These phenomena are quite universal for disordered CIs due to the direct phase transition between the CI and the normal insulator.
\end{abstract}

\date{\today}
\maketitle


\textit{Introduction.---}
Topological band theory plays an essential role in searching for Chern insulators (CIs) \cite{Thouless, Haldane, HMWeng}. A CI is distinguished from a normal band insulator for possessing chiral edge states. Specifically, when the Fermi energy is inside the bulk gap, the electron transport of the chiral edge modes is ballistic, which leads to quantized conductance even for macroscopic samples \cite{CZChang1, CZChang2, XFKou, Checkelsky, YBZhang, YYWang}. Therefore, the CI with a large gap is highly desired for  observing the quantized electron transport. Meanwhile, these dissipationless chiral modes are confined to the boundary of the CI and cannot be manipulated spatially. These two bottlenecks hinder the possible applications of CI in low-power electronic devices \cite{KHe, HMWeng, CXLiu, RYu, ZHQiao, Otrokov}.



In recent years, the CIs have been observed in various systems \cite{CZChang1, CZChang2, XFKou, Checkelsky, YBZhang, YYWang}. However, the CIs' mobility in these experiments are extremely low as from $74\mathrm{cm}^2/(\mathrm{V \cdot s})$ to $760\mathrm{cm}^2/(\mathrm{V \cdot s})$, which suggests the presence of strong disorder. Hence, the celebrated Anderson localization theory may play an important role in these systems \cite{Anderson1, Anderson2, QNiu, Huckestein}.
Naively, the Anderson localization in CI only extends its quantized edge transport region from the band gap to the mobility gap, because the bulk states inside the mobility gap now become localized \cite{DWXu}.
However, such speculation ignores the unique feature of Anderson phase transition in the CIs \cite{Chalker, Mirlin, Beenakker1}.
In principle, the Anderson phase transition is universal and only depends on the system's dimension and its symmetry ensemble \cite{Mirlin, Beenakker1}.
A two-dimension CI breaking time-reversal (TR) symmetry belongs to the unitary ensemble and hence exhibits a direct transition from CI to normal insulator (NI) \cite{Huckestein}. That is to say, all the bulk states in the disordered CI are localized except for the discrete mobility edges \cite{Chalker, Onoda2003, Onoda2007, Yamakage, CZChen}. Consequently, utilizing such unique feature of the Anderson phase transition, we can go beyond the topological band theory and manipulate the transport properties of the CIs so that the bottlenecks above are overcome.

\begin{figure}
\includegraphics[width=7.5cm]{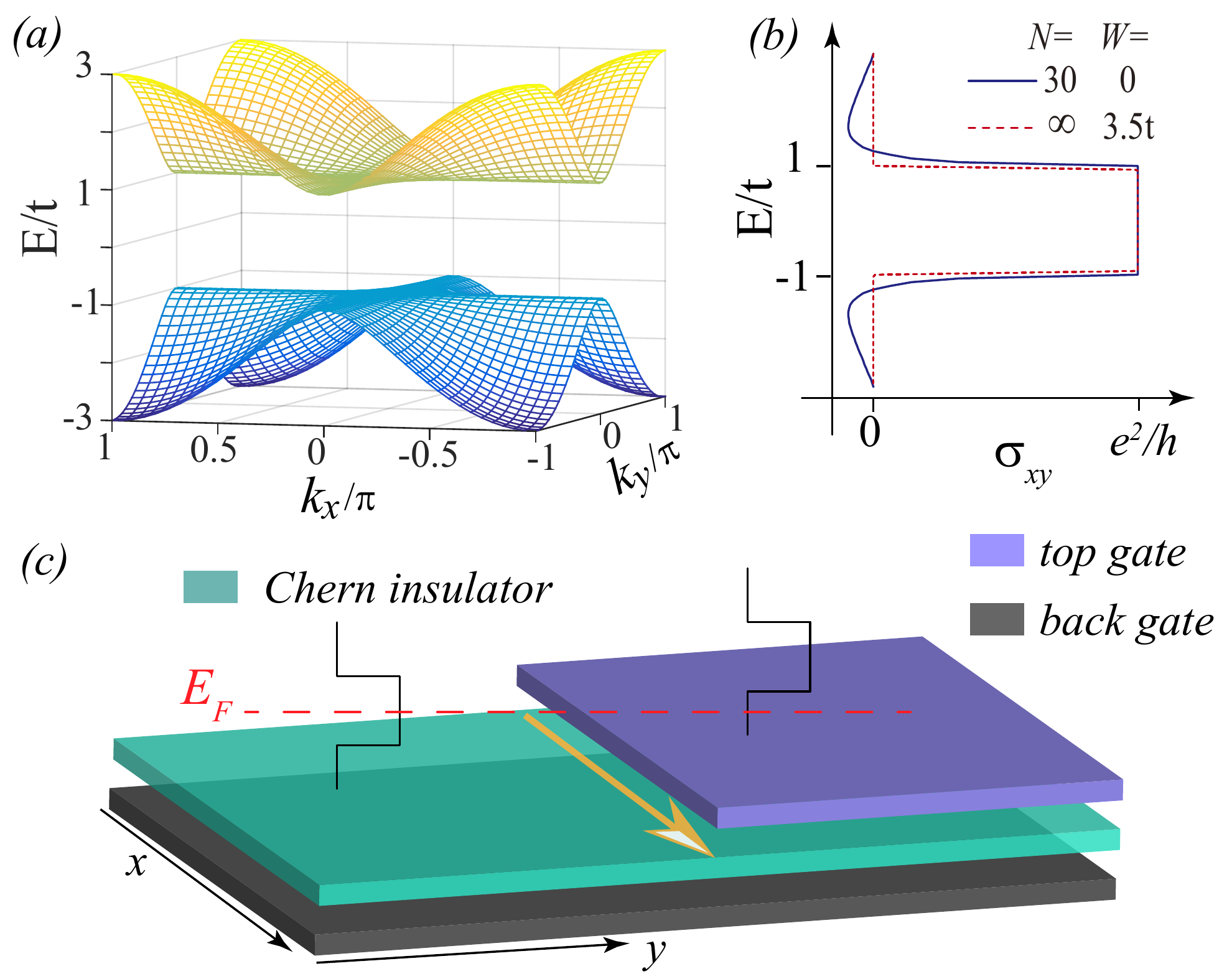}
\caption{(Color online) (a) Typical band spectrum of a CI. (b) Hall conductivity $\sigma_{xy}$ {\it vs} Fermi energy $E$ in clean and disordered samples, respectively. $N$ is the sample size, $W$ is the disorder strength. The red dash line represents $\sigma_{xy}$ in the thermodynamic limit. (c) Schematic plot of the step-potential-induced chiral interface state in a disordered CI.}
\label{Fig1}
\end{figure}

In this Letter, we propose that the chiral interface states with quantized transport can be realized in disordered CIs by combining the Anderson localization and an external step potential.
Figure \ref{Fig1} illustrates the proposed setup and its corresponding physical mechanism.
According to the topological band theory, the Hall conductivity $\sigma_{xy}$ for a clean CI is quantized inside the band gap, while continuously decreases to zero within the conduction band or the valence band by varying the Fermi energy.
In contrast, according to the scaling theory of the disordered CIs \cite{Pruisken}, the insulating feature ensures that $\sigma_{xy}$ jumps from $e^2/h$ to $0$ sharply at two mobility edges since all the bulk states are localized by Anderson localization.
Therefore, when applying a step potential, the Chern numbers (defined as $\sigma_{xy}$ divided by $e^2/h$) of a disordered sample in two separated regions differ by $1$ for specific Fermi energy region [e.g. red dashed line in Fig. \ref{Fig1}(c)].
Therefore, a chiral interface state emerges and the spatial separation of the backscattering channels leads to quantized electronic transport.

From the numerical simulations on the transport behavior for both the Qi-Wu-Zhang (QWZ) model and the Haldane model with a step potential, we verify the existence of the chiral interface states and the quantized transport in the disordered CI. In particular, we find that in addition to the bulk gap, the energy region with quantized transport can also be modulated through the potential applied. Moreover, we also study the disordered CI with a multi-step potential, where the chiral states can be shifted between different interfaces of the potential by varying the Fermi energy. Meanwhile, the energy range for the quantized transport can also be greatly enlarged. Remarkably, such multi-step potential can be replaced by an external electric field along the transverse direction of the sample, which provides an avenue to realize quantized transport with a wide energy window
. Finally, the differences between our proposal and other related topological states are also clarified.


\textit{Theoretical model.---} Both the QWZ model and the Haldane model are widely adopted for investigating the CIs \cite{QWZ, Haldane}. Our investigation is mainly based on the QWZ model whose Hamiltonian in a square lattice reads:

\begin{eqnarray}
\mathcal{H} & = & \sum_{\mathbf{i}} [ c_\mathbf{i}^{\dagger} (\frac{t\sigma_z}{2} -i\upsilon\sigma_y ) c_{\mathbf{i}+\hat{x}} + c_\mathbf{i}^{\dagger} (\frac{t\sigma_z}{2} -i\upsilon\sigma_x) c_{\mathbf{i}+\hat{y}} + h.c. ] \nonumber\\
& & + \sum_{\mathbf{i}} [c_{\mathbf{i}}^{\dagger} (m-2t) \sigma_z c_\mathbf{i} +  c_{\mathbf{i}}^{\dagger} (V_{\mathbf{i}} + W_{\mathbf{i}}) \sigma_0 c_{\mathbf{i}} ].
\end{eqnarray}

\noindent where $c_{\mathbf{i}}^{\dagger}$ is the creation operator on site $\mathbf{i}$ and $\hat{x}$ ($ \hat{y}$) is the unit vector along the $x$ ($y$)-direction, $\sigma_{x,y,z}$ are Pauli matrices and  $\sigma_{0}$ is a $2\times2$ identity matrix. Fermi velocity $\upsilon$, hopping energy $t$ and mass $m$ are three independent Hamiltonian parameters. In a clean QWZ lattice, the band gap and the Chern number is determined by $m$. $V_{\mathbf{i}}$ represents the profile of the applied potential, for example, the top-gate-induced potential in Fig. \ref{Fig1}(c) is in a step function form as $V_{\mathbf{i}} =V \Theta (i_y -N/2)$. Finally, $W_{\mathbf{i}}$ is the Anderson disorder uniformly distributed within $[-\frac{W}{2}, \frac{W}{2}]$ where $W$ indicates the disorder strength.

In addition to the QWZ model, we also study the Haldane model whose Hamiltonian is

\begin{eqnarray}
\mathcal{H}_d = \sum_{\langle \mathbf{i}, \mathbf{j} \rangle} - t c_{\mathbf{i}}^{\dagger}c_{\mathbf{j}} + i \sum_{\langle\langle {\mathbf{i}}, {\mathbf{j}} \rangle\rangle} t_2 c_{\mathbf{i}}^{\dagger} \upsilon_{\mathbf{i,j}} c_{\mathbf{j}} + \sum_{\mathbf{i}} c_{\mathbf{i}}^{\dagger} W_{\mathbf{i}} c_{\mathbf{i}}.
\end{eqnarray}

\noindent in which the three terms are the nearest neighbour direct hopping term, next nearest neighbour spin-orbit coupling (SOC) term, and the Anderson disorder term, respectively \cite{Onoda2003}. In our simulation, the parameters are chosen as $\upsilon =0.5t$, $m=t$, and $t_2=0.3t$.

Numerical methods including non-equilibrium Green's function are adopted to study the transport properties of the disordered CIs \cite{Datta, Kramer, Prodan}. Although the experiments are usually conducted on macroscopic samples \cite{CZChang1, CZChang2, XFKou, Checkelsky, YBZhang, YYWang}, only much smaller samples can be numerically investigated due to the limit of the computing power. Therefore, a finite-size scaling analysis is adopted, in which a series of samples with different sample sizes $N$ are simulated and an extrapolation is performed to obtain the transport properties in the thermodynamic limit ($N \to \infty$).

\begin{figure}
\includegraphics [width=8.0 cm]{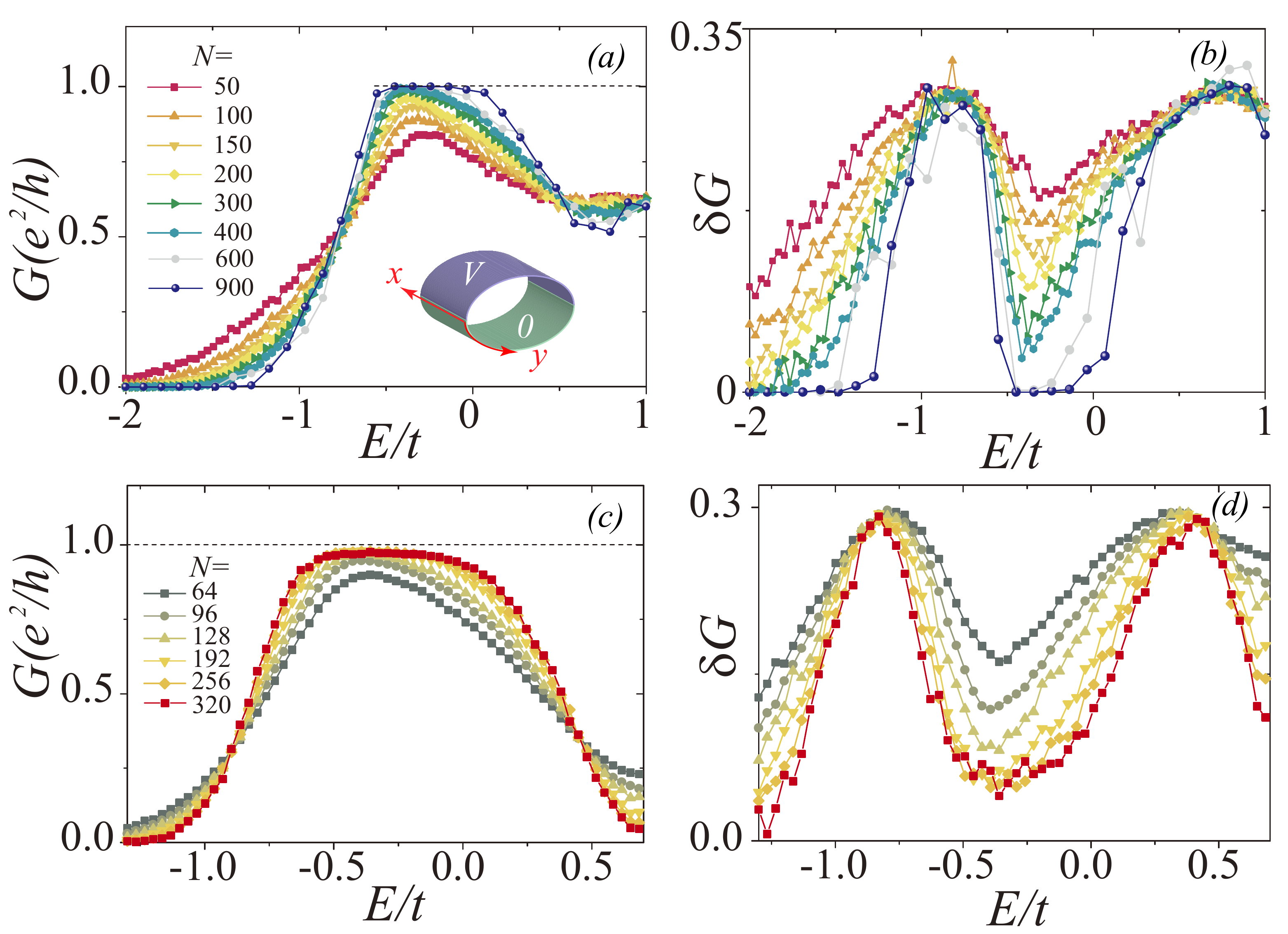}
\caption{(Color online) The differential conductance $G$  and the conductance fluctuation ${\rm \delta G}$ {\it vs} Fermi energy $E$ in a two terminal setup. The current flows along the $+x$ direction.
The step potential $V$ and the disorder is only presented in the central region. Here, up to $1000$ disorder configurations are averaged. 
(a), (b) QWZ model with $V=1.5t$ and disorder strength $W=3.5t$. The sample size is $Na \times Na$, where $a$ is the lattice constant.
(c), (d) Haldane model with $V=1.4t$ and sample size $3Na \times 2\sqrt{3}Na$, other parameters are the same as those in Fig. 3(a) of Ref. \cite{Onoda2007}.}
\label{Fig2}
\end{figure}


\textit{Chiral interface states.---}
We first simulate the electron transport in a two-terminal disordered QWZ lattice which is in a cylinder geometry and imposed with a step potential $V$ [inset of Fig. \ref{Fig2}(a)]. In the $V=0$ case, the zero-temperature differential conductance $G$ vanishes in the gapped region around $E=0$ due to the absence of the edge state \cite{HJiang2009}. On the contrary, when a finite $V$ is applied, a $G$ plateau is formed as Fig. \ref{Fig2}(a).
 By increasing the sample size $N$, the plateau width gradually increases and $G$ approaches the quantized value between two mobility edges with $dG/dN=0$. In the same time, the conductance fluctuation $\delta G$ vanishes with the increase of $N$. The quantized $G$ and vanishing $\delta G$ under strong disorder indicates a chiral state in which its conter-progagating partner is spatially well separated.
It is quite consistent with the prediction of the step-potential-induced chiral interface states [Fig. \ref{Fig1}(c)] \cite{footnote1,bulk}.

To verify the universality of the chiral interface states in disordered CIs, the transport properties of the disordered Haldane model is also studied. Here, we do not specialize the parameters, but choose the SOC strength $t_2$ and the disorder strength $W$ adopted in Fig. 3(a) in Ref. \cite{Onoda2007}. For $V=1.4t$, the same $N$-tendency of both $G$ and $\delta G$ versus $E$ as those in the QWZ model is observed. Indeed, as plotted in Fig. \ref{Fig2}(c), (d), all the $N$-curves cross at both $E_{c1} \approx -0.9t$ and $E_{c2} \approx 0.36t$, manifesting a step jump of $G$ from $e^2/h$ to $0$ at these two points in the thermodynamic limit. Such $G$ behavior can be explained as follows.
In both the two sample regions with applied potentials ``$0$'' and ``$V$'', all the bulk states are localized between two mobility edges $E_{c1}$ and $E_{c2}$ where a direct Anderson transition from CI to Anderson insulator occurs. Therefore, a chiral interface state emerges between $E_{c1}$ and $E_{c2}$ and leads to quantized $G$ when a suitable $V$ is applied. From the transport study on both these two models, we conclude that the chiral interface state and the related quantized transport originates from the universal characters of the Anderson transition in the disordered CIs and is independent of the Hamiltonian details. Furthermore, the quantized $G$ is also insensitive to the smoothness of the step potential \cite{Supp,Sp}.

Now we discuss the energy window for observing the quantized transport. The differential conductance $G$ under different $V$ and $W$ is investigated by numerical calculations (Figs. S2-S6 in the Supplementary Materials \cite{Supp}). Generally, the width of such energy window is equal to $E_{c2}-E_{c1}$. Moreover, $E_{c2}-E_{c1}$ is insensitive to the disorder strength $W$ while approximates to the applied potential $V$.
This is quite different from the normal case that $G$ is only quantized inside the CI's mobility gap. Experimentally,
the mobility gap itself is hard to be modulated. However, in the presence of such chiral interface state, the energy window $E_{c2}-E_{c1}$ could be manipulated by tuning the step potential $V$ to obtain the desired quantized transport in the disordered CIs.

\begin{figure}
\includegraphics[width=8.5cm]{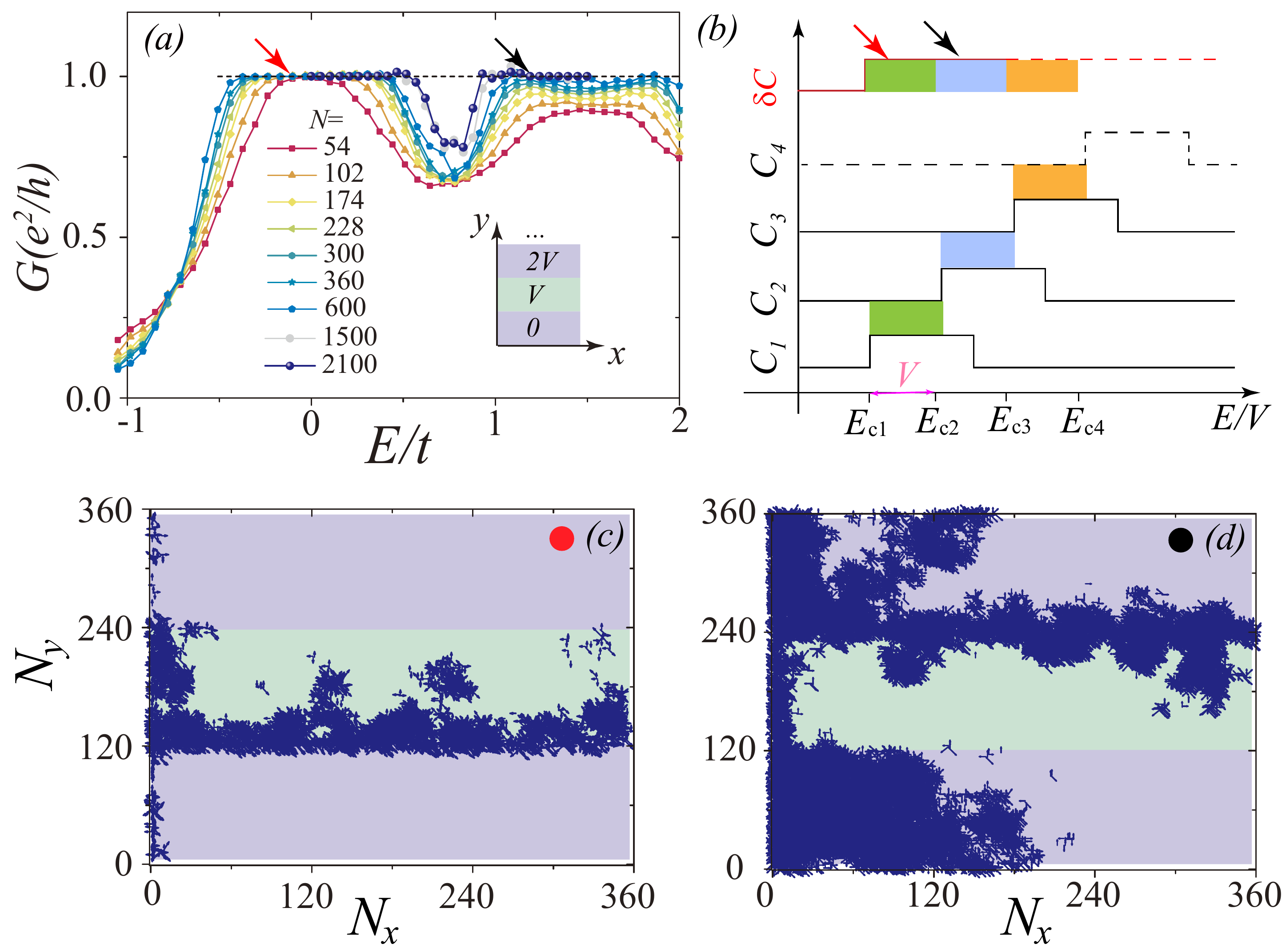}
\caption{(Color online) (a) $G$ {\it vs} Fermi energy $E$ for the samples with multi-step potential. The potential configuration is illustrated in the inset. Here, $V=1.5t$, and $W=3.5t$.
(b) Sketch of the disordered CI's quantized transport energy window with multiple width. Solid lines exhibit the Chern numbers $\mathcal{C}_1$, $\mathcal{C}_2$, $\mathcal{C}_3$ {\it vs} $E$ at the ``$0$'', ``$V$'' and ``$2V$'' labeled regions in the inset of (a), respectively. Chiral interface state emerges in different energy regions (filled with different colors) for different $E$ and leads to a quantized $G$.
(c), (d) Local current density distribution of the sample at energies as marked by the red (black) arrows in (a).}
\label{Fig3}
\end{figure}

Nevertheless, from the sketch in Fig. \ref{Fig1}(c) and the transport results summarized in the Supplementary Materials \cite{Supp}, one can intuitively obtain the consequence that in the presence of a single step potential, such quantized transport energy window tuned by $V$ will not be wider than the mobility gap. Currently, the biggest challenge hinder the application of the CI is its narrow energy gap (narrow quantized region). To overcome such a problem, we further study the transport properties of the disordered CI in the presence of a multi-step potential.
Fig. \ref{Fig3}(a) shows the two-terminal differential conductance $G$ under potential $V_{\mathbf{i}} =V \Theta (i_y -N/3) + V \Theta (i_y -2N/3)$ [inset of Fig. \ref{Fig3}(a)], in which two separated $G$ plateaus at small $N$ are exhibited. As $N$ approaches the thermodynamic limit, these two plateaus will merge into one with perfectly quantized $G=e^2/h$. In this way, significantly, the energy window for quantized $G$ is approximately doubled to $2V=3t$, which is much larger than the original energy gap $2t$ in the clean CI.
For regions labeled by ``$0$'', ``$V$'', and ``$2V$'' in the inset of Fig. \ref{Fig3}(a), due to the Anderson-disorder-induced direct NI-CI transition, their Chern numbers $\mathcal{C}_1$, $\mathcal{C}_2$, $\mathcal{C}_3$ all exhibit a $0 \to 1$ step transition [Fig. \ref{Fig3}(b)] at energies $E_{c1}$, $E_{c2} \approx E_{c1}+V$, and $E_{c3} \approx E_{c2}+V$, respectively. For Fermi energy $E \in (E_{c1}, E_{c2})$, $\mathcal{C}_1-\mathcal{C}_2=1$ so that a chiral interface state emerges along the interface between region ``$0$'' and region ``$V$''. Similarly, for $E \in (E_{c2}, E_{c3})$, the chiral interface state shifts to the interface between region ``$V$'' and region ``$2V$'' since $\mathcal{C}_2-\mathcal{C}_3=1$.
Such physics picture is directly proved by the local current density calculation. When $E$ takes the value as that labeled by the red arrow in Fig. \ref{Fig3}(a), the local current flows along the interface between regions ``$0$'' and ``$V$'' [Fig. \ref{Fig3}(c)]. On the contrary, if $E$ is in the value labeled by the black arrow in Fig. \ref{Fig3}(a), the local current distributes along the interface between regions ``$V$'' and ``$2V$'' [Fig. \ref{Fig3}(d)]. Moreover, in both Fig. \ref{Fig3}(c), (d), no backscattered local current is observed. Finally, one can also adopt a potential profile with more steps to obtain a quantized $G=e^2/h$ region with multiple width, which is illustrated by the $\delta C$ curve filled with different colors in Fig. \ref{Fig3}(b).

Combining the numerical results for both the single-step and the multi-step potential, we claim that with the help of the Anderson localization, the quantized transport energy region can be manipulated from nearly zero to multiple times of the CI's bulk gap. In other words, a continuously tunable transport gap can be obtained in the disordered CIs based on the chiral interface states.

\begin{figure}
\includegraphics [width=8.5 cm]{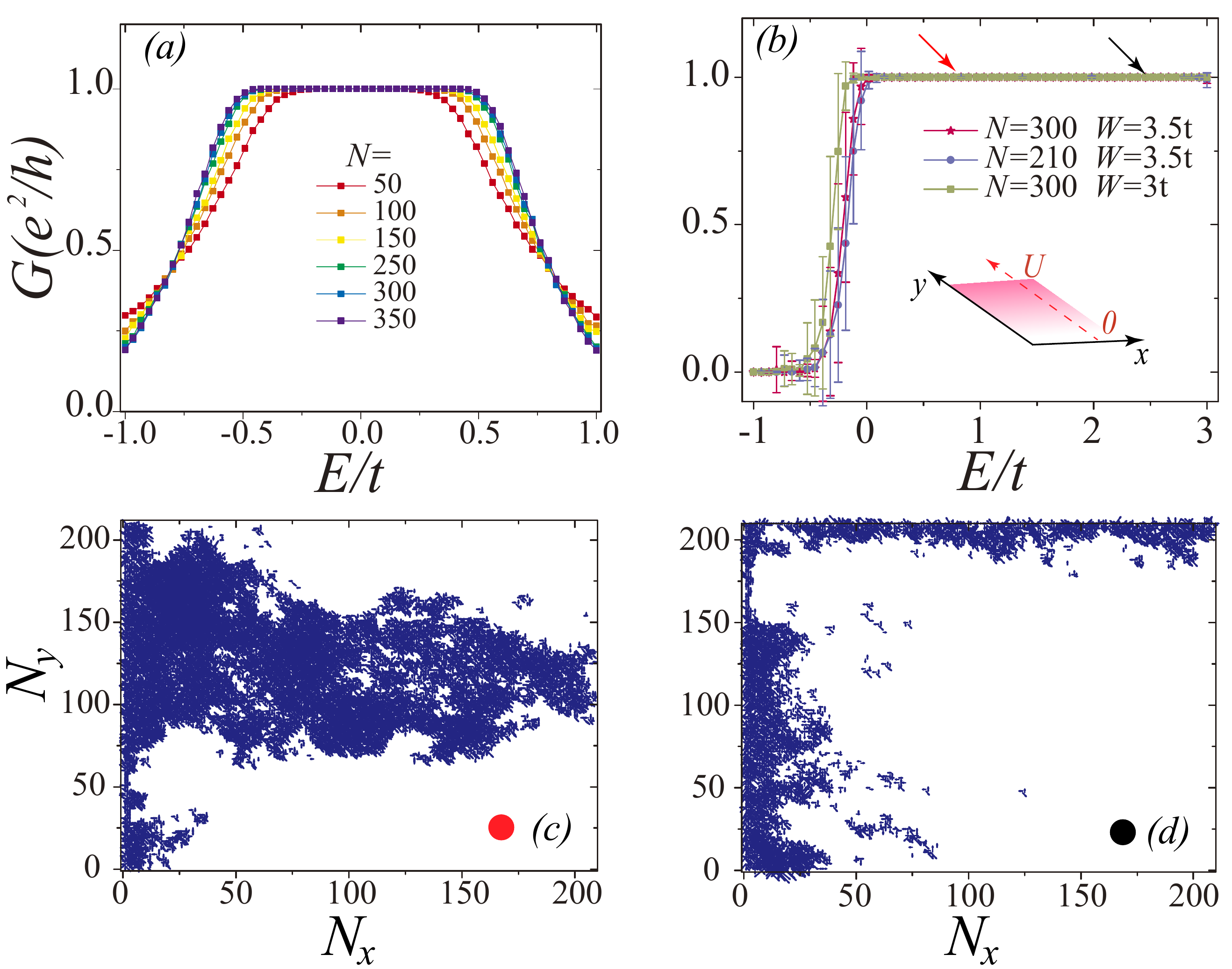}
\caption{(Color online) (a), (b)  $G$ {\it vs}  $E$ for QWZ lattices in the (a) absence; (b) presence of the transverse electric field $E_y$, respectively. Open boundary condition is adopted here and $E_y$ induces a linear potential $V_{\mathbf{i}}$ along the $y$ direction where the maximum potential difference is $U=3t$. The error bar in (b) represents $\delta G$.
(c), (d) Local current density distribution of the sample at the energy marked by the red and black arrows in (b), respectively.  }
\label{Fig4}
\end{figure}


\textit{Experimental proposal.---}
Now we put forward a more realistic experimental proposal.
Although multi-step potential can in principle overcome CIs' narrow gap problem, fabricating such a multi-step potential could be experimentally difficult. As illustrated in Fig. \ref{Fig3}(b), for step potential with arbitrary number of steps, there is always only one conducting chiral interface state in the whole lattice, although its spatial position depends on the explicit value of $E$.  Such results imply that the profile of the applied potential can be relaxed to a monotonously increasing potential along the transverse direction, which can be approximated by an infinite-step potential profile.

We then apply an electric field $E_y$ to the disordered CI lattice, which introduces a linear potential along the transverse direction as shown in the inset of Fig. \ref{Fig4}(b). The maximum potential difference is $U=3t$, which is the same as that in Fig. \ref{Fig3}(a). For comparison, we also study the disordered CI sample in the absence of $E_y$. To directly simulate the experimental observable, open boundary condition is adopted for both these two samples.
In the absence of both disorder and $E_y$, the CI possesses an intrinsic band gap with width $2t$. In the presence of strong Anderson disorder [Fig. \ref{Fig4}(a)], the differential conductance $G$ is quantized when $E \in [-0.82t, 0.82t]$, which indicates the mobility gap for the disordered CI is about $1.64t$ and smaller than the band gap.
In sharp contrast, if $E_y$ is applied, the lattice has no global gap in the clean limit (Fig. S7(d) in the Supplementary Materials \cite{Supp}), while shows a quantized $G$ without fluctuation for $E \in [0, 3t]$ when disorder is presented [Fig. \ref{Fig4}(b)]. Such a $3t$-wide quantized transport region, much broader than the mobility gap $1.64t$ and the band gap $2t$, originates from the potential-induced chiral interface states. The presence of the chiral interface state under strong disorder is verified by the local current distribution [Fig. \ref{Fig4}(c)]. With the increase of $E$, the chiral interface state evolves into the chiral edge state [Fig. \ref{Fig4}(d)]. Moreover, we find that the width of the quantized energy region is proportional to the maximum potential difference $U$, while independent of the disorder strength $W$ and the topological mass $m$ \cite{Supp}. In principle, a large enough potential slope $\frac{U}{N}$ may ruin the quantization of $G$. However, in realistic samples, the quantized region has been greatly enhanced before such slope is large enough \cite{Supp}.

Very recently, the existence of Anderson localization in CI has been suggested by an experiment \cite{YBZhang} in MnBi$_2$Te$_4$ \cite{Supp}. For strong magnetic field, the authors observed the coexistence of the quantum Hall (QH) effect and the quantum anomalous Hall effect, which indicates that the Fermi energy is inside the bulk band.
The QH effect fades away when the magnetic field decreases, however, a quantized Hall conductance is still observed, which strongly implies the localization of the bulk states.
As shown in Fig. \ref{Fig4}(a), the Anderson localization itself has little effect on the narrow gap problem of CI. 
Nevertheless, we expect that applying a transverse electric field in these Anderson-localization-dominated CIs \cite{CZChang1, CZChang2, XFKou, Checkelsky, YBZhang, YYWang} will greatly enlarge their energy window for dissipationless transport.


\textit{Discussion and conclusion.---}
The Anderson phase transition is beyond the scope of the band theory. Such feature distinguishes our proposal from another well studied disorder-induced topological state --- topological Anderson insulator (TAI) \cite{HJiang2009, JianLi, HMGuo, Hughes, Rechtsman,LC}. The TAI originates from the disorder-renormalized band structure \cite{Beenakker2}.
Furthermore, due to the symmetry-related classification of the Anderson phase transition, our main results cannot be extended to metals or the TR-symmetry-protected topological systems \cite{SEVK}, e.g. quantum spin Hall insulator.
 Specifically, due to the symplectic ensemble classification of the latter system, a metallic phase emerges between the topological insulator and the NI phases \cite{Onoda2007, Yamakage, CZChen}. Such metallic phase prohibits the integer topological invariant difference between the two sides of the potential step so that the topological interface states are now absent. Lastly, though all the investigation here is performed at zero temperature, such enlarged quantized region also benefits the observation of the quantized transport in the finite temperature case. Meanwhile, the finite temperature will reduce the carriers' quantum coherence, which is critical to the Anderson localization \cite{Anderson1}. 

In summary, we find a step potential will give rise to the chiral interface state in the disordered CIs. Such state is highly related to the universal Anderson transition properties of the CIs that the disorder drives a direct transition from CI to NI.
Based on such chiral interface state, it is hopeful to breakthrough the bottleneck of the CIs' narrow gap problem and obtain a quantized transport whose energy window could be tuned from nearly zero to multiple times of the band gap. Finally, utilizing a simple transverse electric field, these proposals can be realized under current experimental techniques.

\textit{Acknowledgements.---} We are grateful to H. W. Liu, X. Dai, K. Chang, J. S. Zhang,  and X. G. Wan for helpful discussion. This work was supported by National Basic Research Program of China (Grants No. 2019YFA0308403 and No. 2017YFA0303301), NSFC under Grants Nos. 11534001, 11822407, and 11874274, and a Project Funded by the Priority Academic Program Development of Jiangsu Higher Education Institutions (PAPD). J. Liu acknowledges supports from the RGC (N\_HKUST626/18,26302118 and 16305019). Z. Q. Zhang and C. Z. Chen contribute equally to this work.

\begin{widetext}
\newpage


\setcounter{equation}{0}
\setcounter{figure}{0}
\setcounter{table}{0}
\renewcommand{\theequation}{S\arabic{equation}}
\renewcommand{\thefigure}{S\arabic{figure}}
\renewcommand{\bibnumfmt}[1]{[S#1]}
\renewcommand{\citenumfont}[1]{S#1}

\begin{center}
\textbf{Supplementary Materials for ``Chiral interface states and related quantized transport in disordered Chern insulators"}
\end{center}

\begin{center}
Zhi-Qiang Zhang$^{1}$, Chui-Zhen Chen$^{2,1}$, Yijia Wu$^{3}$, Hua Jiang$^{1,2,*}$, Junwei Liu$^{4}$,\\ Qing-Feng Sun$^{3,5,6}$, and X. C. Xie$^{3,5,6}$
\end{center}

\begin{center}
$^1$~{\it School of Physical Science and Technology, Soochow University, Suzhou 215006, China}

$^2$~{\it Institute for Advanced Study, Soochow University, Suzhou 215006, China}

$^3$~{\it International Center for Quantum Materials, School of Physics, Peking University, Beijing 100871, China}

$^4$~{\it Department of Physics, Hong Kong University of Science and Technology, Kowloon, Hong Kong}

$^5$~{\it Beijing Academy of Quantum Information Sciences, Beijing 100193, China}

$^6$~{\it CAS Center for Excellence in Topological Quantum Computation, University of Chinese Academy of Sciences, Beijing 100190, China}
\end{center}

\tableofcontents

\section{S1. Finite size scaling of the Hall conductance}

\begin{figure}[t]
\includegraphics[width=8cm]{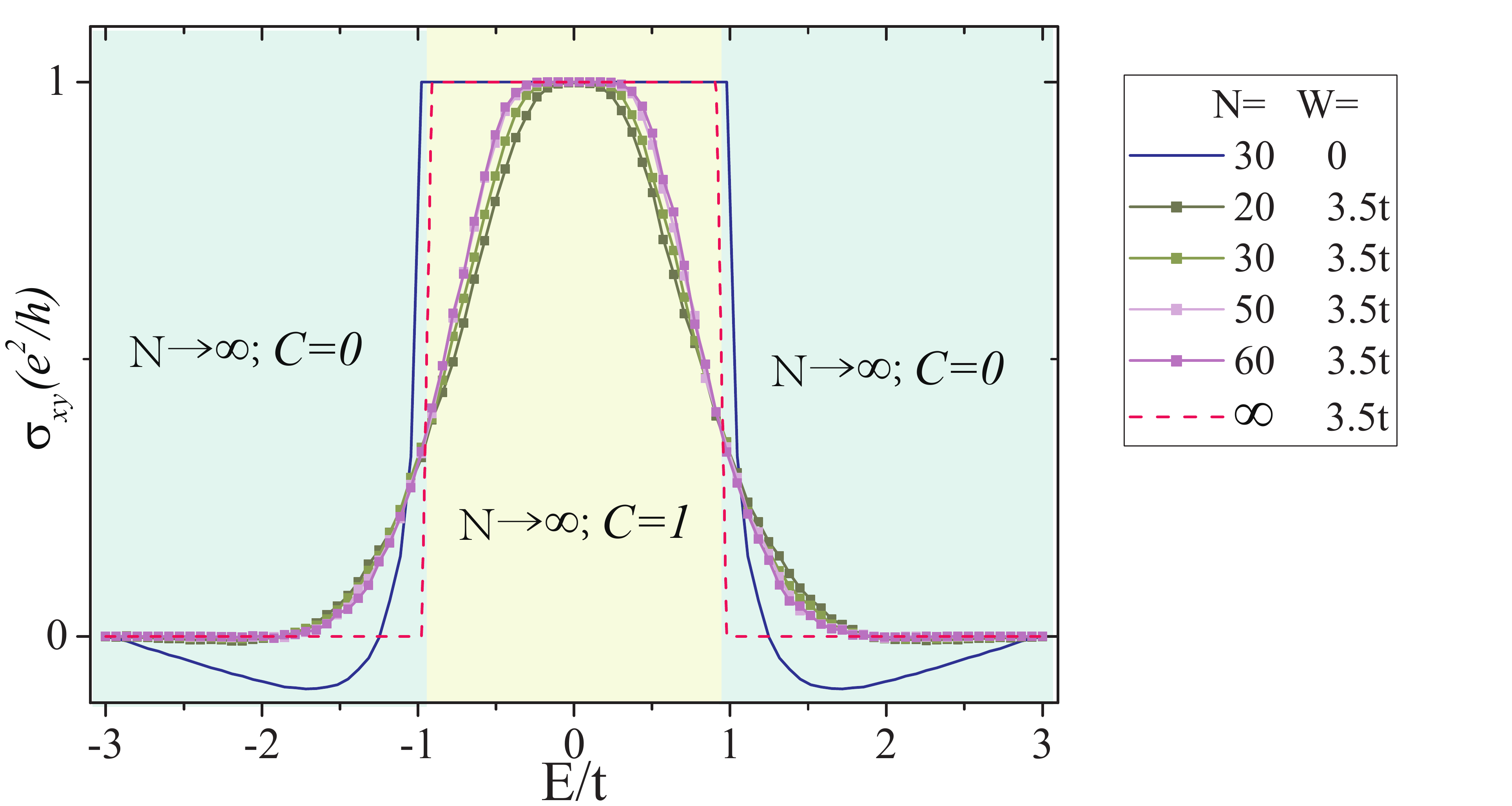}
\caption{The Hall conductance $\sigma_{xy}$ versus the Fermi energy $E$ for different sample sizes $N$ and different disorder strengths $W$. All these curves are calculated by using the QWZ model. Other parameters are the same as those in the main text.
}\label{S0}
\end{figure}

To show that the Hall conductance $\sigma_{xy}$ trace is in the shape of the red dash line in Fig. 1(b) of the main text (or the red dash line in Fig. \ref{S0}) in the thermodynamic limit ($N\rightarrow \infty$), we present the scaling analysis of $\sigma_{xy}$ in this section. The Hall conductance is calculated by using the real space Chern number method \cite{sNGM,sNGM1} where

\begin{equation}
\sigma_{xy} = \frac{2\pi ie^2}{h} \mathrm{Tr}[P[-i[x,P],-i[y,P]]].
\label{Chernnumber}
\end{equation}

\noindent Here, $P$ is the projector of the occupied states, and $(x, y)$ is the real space coordinate for each site. The evolution of the Hall conductance $\sigma_{xy}$ with the variance of the Fermi energy $E$ is plotted in Fig. \ref{S0}, in which the calculation is based on the QWZ model \cite{sBHZ} adopted in the main text. We have to point out that Eq. \ref{Chernnumber} is also corresponding to the Chern number with $C=\frac{h}{e^2}\sigma_{xy}$, which has been presented in Ref. \cite{sNGM1}.

In case of $W=0$, $\sigma_{xy}$ decreases to zero smoothly and there is no sudden step jump between the $\sigma_{xy}=0$ plateau and the $\sigma_{xy}=e^2/h$ plateau. However, after taking the disorder effect into consideration ($W=3.5t$), it can be seen that $\sigma_{xy}$ increases with the increasing of the sample size $N$ when the Fermi energy lies between two critical points. In the meantime, $\sigma_{xy}$ decreases with the increasing of $N$ for the rest part of the Fermi energy. Therefore, it is reasonable to assume that in the thermodynamic limit $N\rightarrow \infty$, the Hall trace will be in the form of a step function as the red dashed line in Fig. \ref{S0}.
To sum up, $\sigma_{xy}=0$ and $\sigma_{xy}=e^2/h$ regions are separated by critical points in the disordered Chern insulators (CIs).


\section{S2. An analysis on the two-terminal conductance}

In this section, we investigate the transport properties of square CI samples under periodic boundary condition in details. In our calculation, the current flows along the $x$ direction, while the electric potential applied along the $y$ direction can be written as $V_{\mathbf{i}}=V\Theta (i_y -N/2)$. Here $\Theta(i_y-N/2)$ is the Heaviside step function, $V$ is the voltage difference, $i_y$ is the real space coordinate along the $y$ direction, and $N$ is the sample width.

\begin{figure}[b]
\includegraphics[width=10cm]{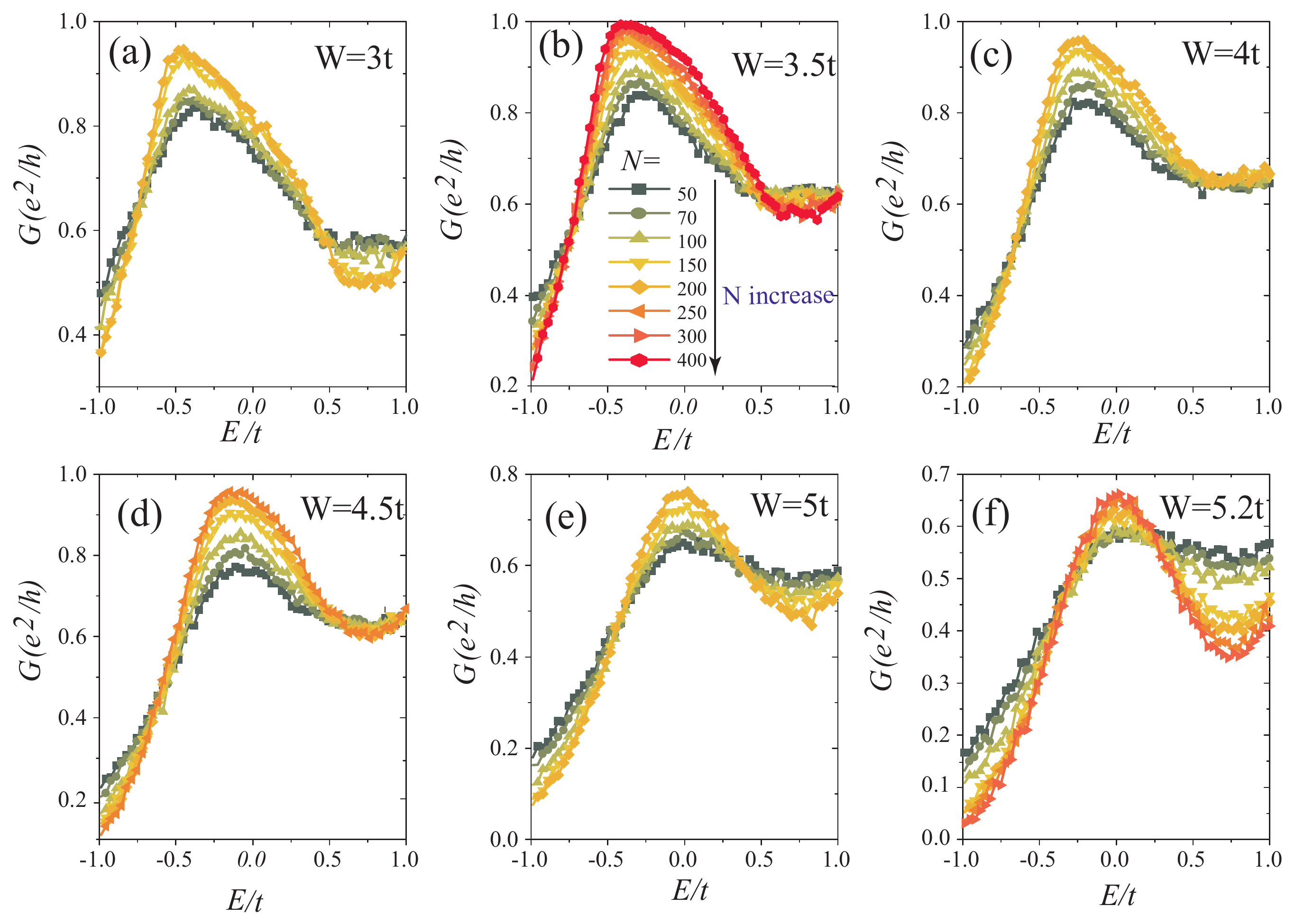}
\caption{The two-terminal conductance along the $x$ direction for the cylindrical CI samples with different disorder strengths $W$ and different sample sizes $N_x=N_y=N$. The electric voltage applied along the $y$ direction can be written as $V_{\mathbf{i}}=V\Theta (i_y -N/2)$ where $\Theta (i_y -N/2)$ is the Heaviside step function, $V$ is the voltage difference ($V=1.5t$ in this figure), and $i_y$ is the real space coordinate along the $y$ direction.
The disorder strength for each figure is (a) $W=3t$; (b) $W=3.5t$; (c) $W=4t$; (d) W=4.5t; (e) $W=5t$; and (f) $W=5.2t$. Each curve is averaged for $1000$ disorder configurations. Other parameters are the same as those in Fig. \ref{S0}.
}\label{S1}
\end{figure}

First, we set $V=1.5t$ and study the relation between the two-terminal conductance $G$ and the disorder strength $W$. Based on the finite-size scaling analysis, the conductance in the thermodynamic limit could be obtained.
To be specific, if $G$ increases (decreases) with the increasing of $N$ (the conductance fluctuation $\delta G$ decreases at the same time), the conductance will approach $e^2/h$ ($0$) in the thermodynamic limit $N\rightarrow \infty$. In contrast, the conductance is nearly independent of the sample size at the transition points. Fig. \ref{S1} exhibits the conductance curves for different sample sizes $N$.
The transition points drawn from Fig. \ref{S1} is summarized in the phase diagram shown in Fig. \ref{S3}(a), in which the conductance approaches $e^2/h$ in the thermodynamic limit ($N\rightarrow \infty$) for the orange region. For simplicity, the low energy transition point is denoted as the ``first transition point'' here, and the high energy transition point is named as the ``second transition point''.
When $W\leq 4t$, both the first and the second transition points shift to the higher energy with the increasing of $W$, and the width of the orange region is almost invariant. On the contrary, for $W>4t$, the width of the non-trivial region quickly decreases with the increasing of $W$.

\begin{figure}
\includegraphics[width=10cm]{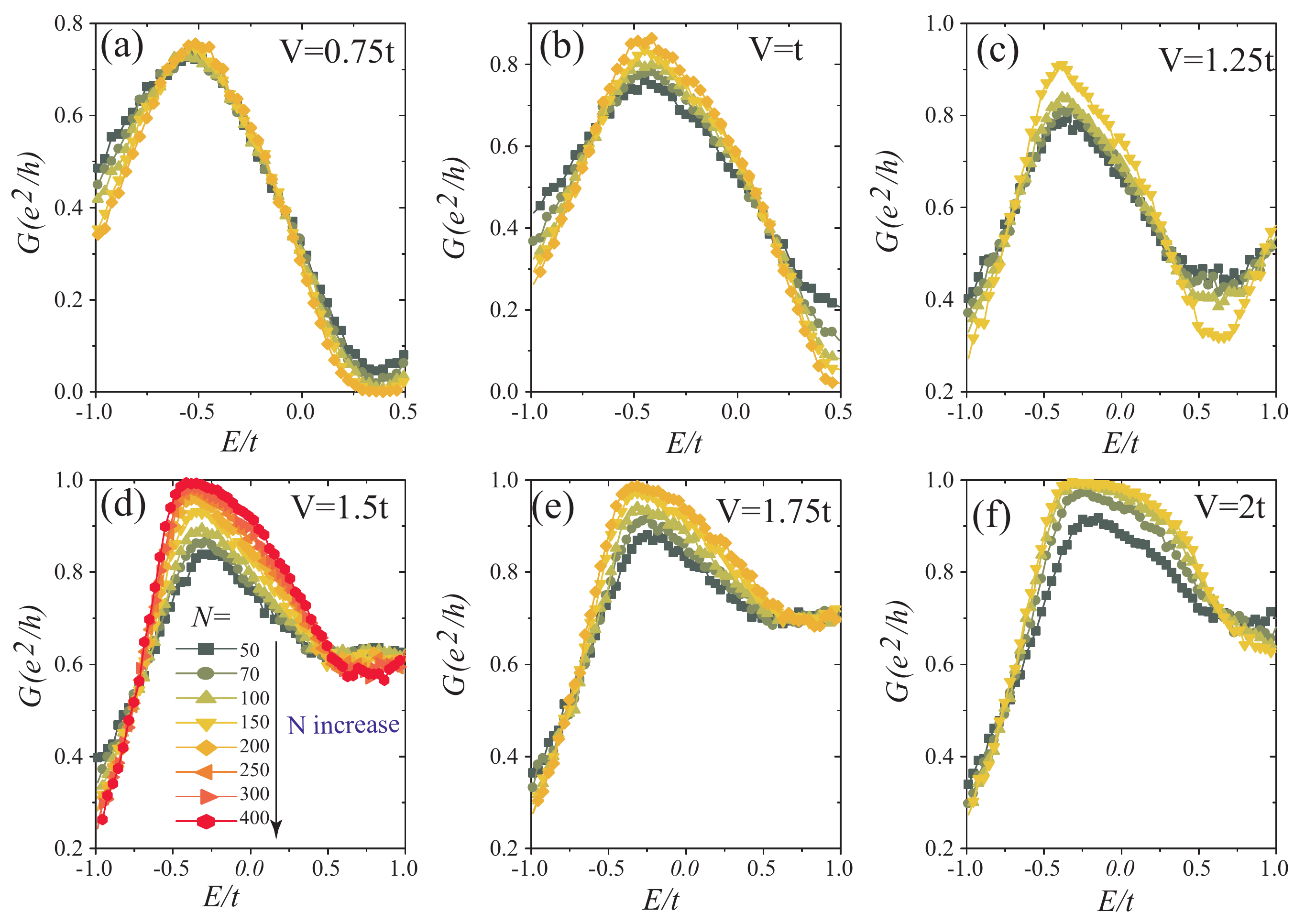}
\caption{The two-terminal conductance along the $x$ direction for the cylindrical CI samples with different voltage differences $V$ and different sample sizes $N$. Disorder strength is fixed at $W=3.5t$ in this figure.
The electric voltage applied along the $y$ direction is the same as that in Fig. \ref{S1}. The voltage difference $V$ for each figure is (a) $V=0.75t$; (b) $V=t$; (c) $V=1.25t$; (d) $V=1.5t$; (e) $V=1.75t$; and (f) $V=2t$. Each curve is averaged for $1000$ disorder configurations. Other parameters are the same as those in Fig. \ref{S0}.
}\label{S2}
\end{figure}

As we have shown in Fig. 1 in the main text, the non-trivial region shown in Fig. \ref{S3} is closely related to the chiral interface states.
To illustrate such topologically non-trivial region in Fig. \ref{S3}(a), we exhibit the evolution of the Chern number at both the ``0'' and the ``V'' regions for different disorder strengths as shown in Fig. \ref{S3}(c).
The solid red (black) line indicates the Chern number as a function of the Fermi energy $E$ at the ``V'' (`0'') region, respectively. The inset of Fig. \ref{S3}(c) plots the voltage distribution in real space, where the ``0'' (``V'') region corresponds to the green (purple) part of the CI sample.
The green dashed line labels the Fermi energy $E=0$ in Fig. \ref{S3}(a), while the red dashed line indicates the Fermi energy $E=V$. If the Chern numbers shown by the solid black and the solid red lines are equal to $1$ and $0$ at the same energy, respectively, then the chiral interface state protected by the Chern number difference is presented [Fig. \ref{S3}(c)].
Moreover, the topological band gaps for both the solid red line and the solid black line decrease with the increasing of the disorder strength $W$.

\begin{figure}[t]
\includegraphics[width=11cm]{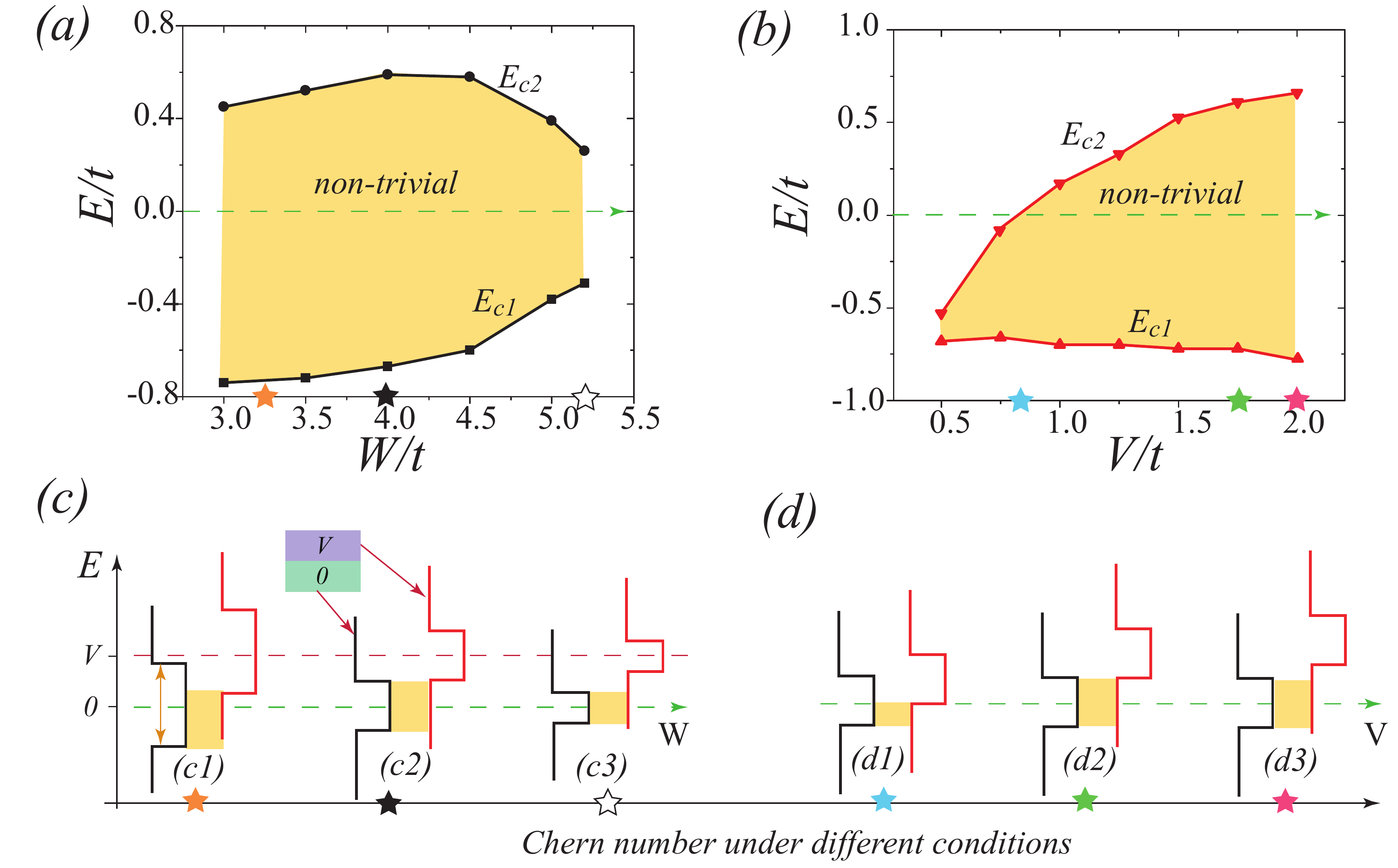}
\caption{Phase diagrams obtained from Fig. \ref{S1} and Fig. \ref{S2}. The conductance in the orange area is $e^2/h$ in the thermodynamic limit ($N\rightarrow \infty$).
(a) Phase diagram on the plane spanned by the Fermi energy $E$ and the disorder strength $W$.
(b) Phase diagram on the plane spanned by the Fermi energy $E$ and the voltage difference $V$.
(c) The Chern number versus the Fermi energy $E$ for different disorder strengths $W$. The solid red (black) line is the Chern number curve for the sample area with electric voltage $V$ ($0$). The inset is a sketch of the voltage distribution in real space.
The green dashed line corresponds to the $E=0$ green dashed line in (a). The red dashed line indicates the Fermi energy $E=V$.
(d) The Chern number versus the Fermi energy $E$ for different voltages $V$. Other parameters are the same as those in (c).
}\label{S3}
\end{figure}

Now we explain the phase boundaries of the non-trivial region in Fig. \ref{S3}(a).
First, for weak disorder strength ($W<4t$), the lower (upper) phase boundary in Fig. \ref{S3}(a) is determined by the lower edge of the solid black (red) line [Fig. \ref{S3}(c1)].
It can be seen that both the lower edges of the solid black  line and the solid red line shift to the higher energy with the increasing of the disorder strength $W$. In this way, the width of the orange region only depends on $V$ and is almost independent of the disorder strength $W$ in the case of $W<4t$.
Then, for $W\approx 4t$, the lower edge of the solid red line begins to be higher than the upper edge of the solid black line [Fig. \ref{S3}(c2)]. Therefore, finally, the orange region corresponds to the topological gap of the solid black line in case of $W>4t$ [Fig. \ref{S3}(c3)]. Hence the width of the orange area decreases with the increasing of $W$. We also notice that for $W>5.5t$, the non-trivial region disappears since the topological state is destroyed by disorder.

\begin{figure}[b]
\includegraphics[width=6.5cm]{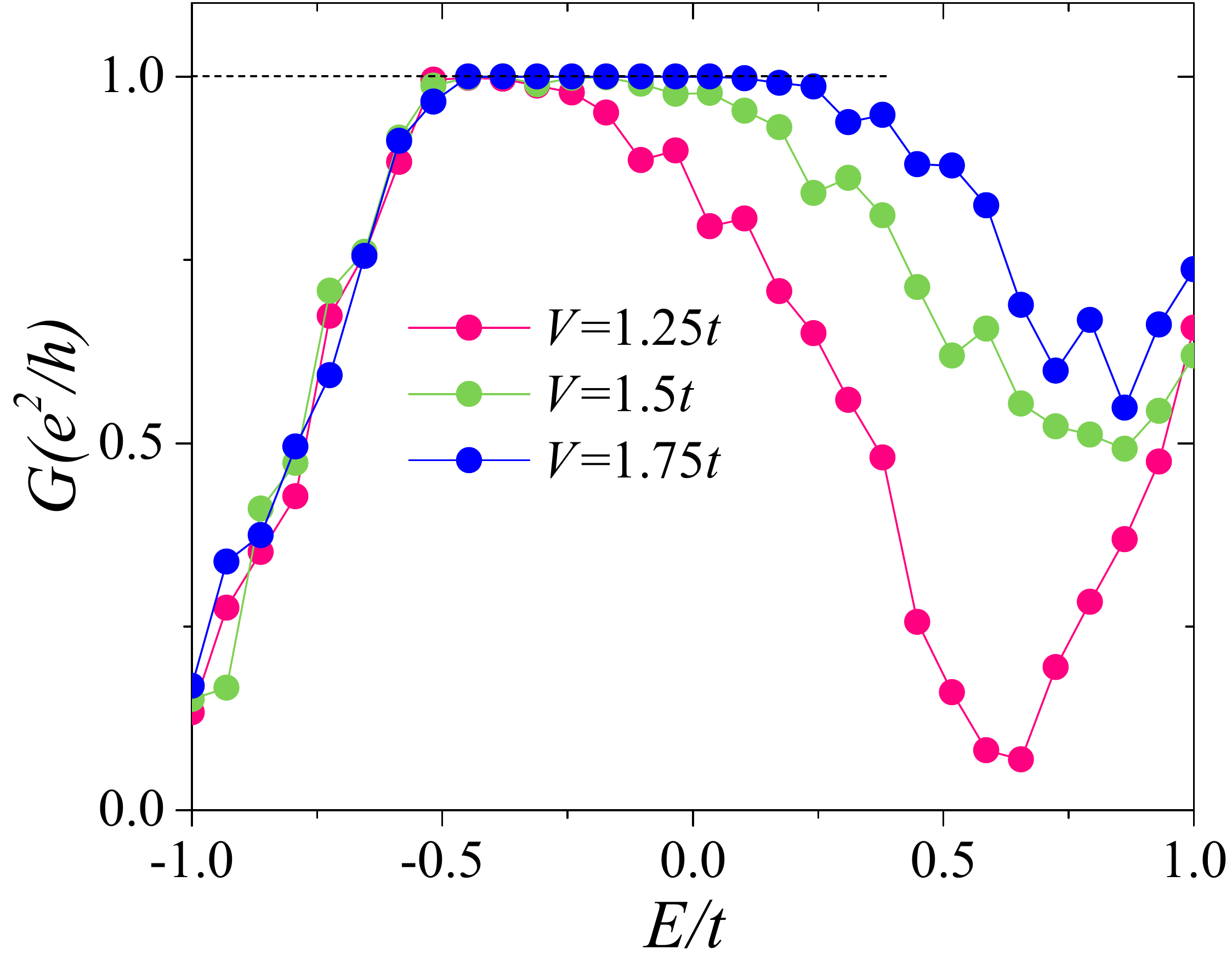}
\caption{The evolution of the two-terminal conductance $G$ with respect to the Fermi energy $E$ for different voltage differences $V$. The disorder strength is chosen as $W=3.5t$, and the sample size is $N=900$. Other parameters are the same as those in Fig. \ref{S1}.
}\label{S4}
\end{figure}

To further investigate the dependence of the topological region on the gate voltage, we study the two-terminal conductance $G$ as a function of the voltage difference $V$ with $W=3.5t$. The conductance traces are shown in Fig. \ref{S2}, and the corresponding transition points are summarized as Fig. \ref{S3}(b). The first transition point is almost invariant with the increasing of $V$, while the second transition point strongly depends on $V$. Moreover, it is shown that the gradient of the non-trival region's upper boundary (formed by the second transition points) decreases with the increasing of $V$. Such results is also illustrated in Fig. \ref{S3}(d).
Since the first transition point is determined by the lower edge of the solid black line ($V$-independent), the lower boundary of the non-trivial region is nearly independent of $V$. The slight shift of the first transition point could originate from the self-energy of the purple area [the inset of Fig. \ref{S3}(c)], which is $V$-dependent.
The second transition point is sensitive to $V$ because these transition points are determined by the lower edge of the solid red line, which corresponds to the purple sample region imposed with electric voltage $V$. Furthermore, since the width of the orange region is almost invariant when $V$ is larger than the topological band gap of the solid black line [Fig. \ref{S3}(d2), (d3)], the gradient of the topologically non-trivial region's upper boundary in Fig. \ref{S3}(b) decreases as expected. These results imply that it is an efficient way modulating the topological ``band gap'' by tuning the gate voltage $V$. Although such single-step potential has a disadvantage that the topological ``band gap'' cannot be wider than the mobility gap at the gate voltage $V=0$, such disadvantage could be overcame by using the multi-step electric potential proposed in the main text.

Finally, as shown in Fig. \ref{S4}, in case of fixed sample size $N=900$ and fixed disorder strength $W=3.5t$, the quantized conductance $G=e^2/h$ is much easier to be obtained for larger $V$. Besides, the width of the $G= e^2/h$ plateau also increases with the increasing of $V$. This implies that tuning the gate voltage is also an efficient way modulating the quantized plateau in disordered CIs.


\section{S3. Two-terminal conductance for linear biased Chern insulator samples}

In the section, we study the two-terminal conductance $G$ in the linear biased CI samples, in which a linear transverse voltage is imposed along the $y$ direction as $V_\mathbf{i} = i_y V/N$ ($i_y$ is in the range of $[1, N]$).
The conductance $G$ as a function of the Fermi energy $E$ as well as the voltage difference $V$ is plotted in Fig. \ref{S5}. As shown in Fig. \ref{S5}(a), the width of the plateau with quantized conductance $G=e^2/h$ increases almost linearly with the increasing of $V$.
The data in Fig. \ref{S5}(a) could be replotted as Fig. \ref{S5}(b) to obtain a more clear view, where the $G=e^2/h$ plateau is evidently broadened for larger $V$. These results suggest that the topological ``band gap'' can be greatly enlarged with the help of the transverse voltage along the $y$ direction.

\begin{figure}[t]
\includegraphics[width=14cm]{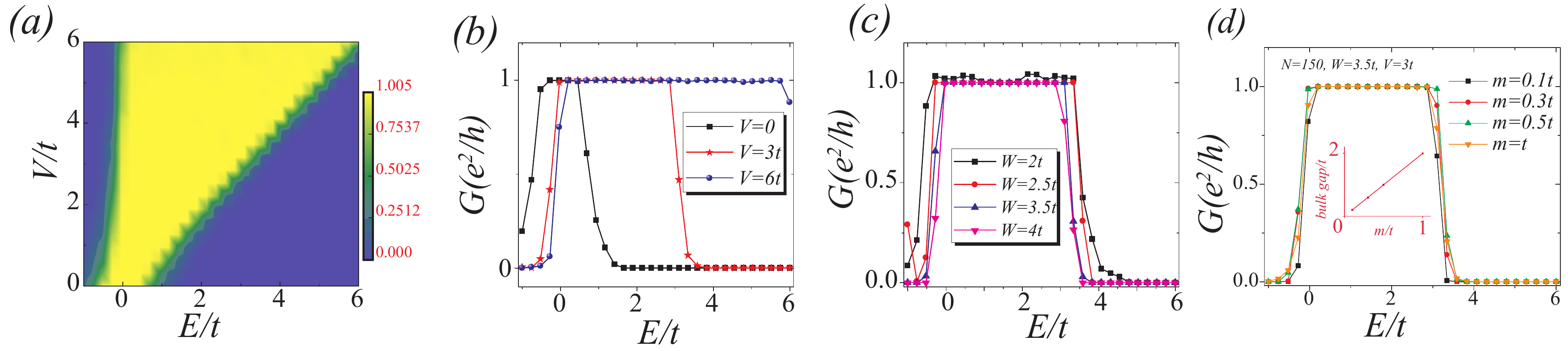}
\caption{(a) The two-terminal conductance $G$ versus the voltage difference $V$ and the Fermi energy $E$ for rectangle CI samples in the size of $N_x=2N$, $N_y=N$ ($N=300$), and disorder strength $W=3.5t$. The voltage applied along the $y$ direction is in the form of a linear function as $V_\mathbf{i} = i_y V/N$, where $i_y$ is in the range of $[1, N]$. Other parameters are the same as those in Fig. \ref{S0}.
(b) The conductance $G$ versus the Fermi energy $E$ for different voltage differences $V$.
(c) The conductance $G$ versus the Fermi energy $E$ for different disorder strength $W$ with $N=600$ and $V=3t$.
 (d) $G$ versus $E$ under different $m$. As shown in the inset, $m$ determines the bulk gap of the clean samples in the absence of external voltage.
}\label{S5}
\end{figure}

Fig. \ref{S5}(c) shows the conductance traces for different disorder strengths $W$ with sample size $N=600$. It implies that the conductance plateau width is nearly independent of the disorder strength $W$, and such plateau can be formed in a wide range of disorder strength $W\in[2.5t,~4t]$.
We also notice that the conductance slightly deviates from the quantized value $G=e^2/h$ in the case of $W=2t$. Such deviation originates from the small disorder strength and the small sample size in which some bulk states can still contribute to the conductance $G$. On the contrary, for a CI sample with larger sample size, the conductance can be perfectly quantized at $G=e^2/h$.

Moreover, as shown in Fig. \ref{S5}(d), even in the small-gap case such as $m=0.1t$, the quantized plateau is still well preserved and the plateau width is also almost unchanged. Therefore, the main conclusion of our manuscript is still valid for the small-gap case, since the width of the energy window with quantized transport is almost independent of the bulk gap of the clean sample.


\section{S4. The band structures in different voltage schemes}

In this section, we discuss the band structures of the QWZ lattice in four different cases.

\begin{figure}[t]
\includegraphics[width=15cm]{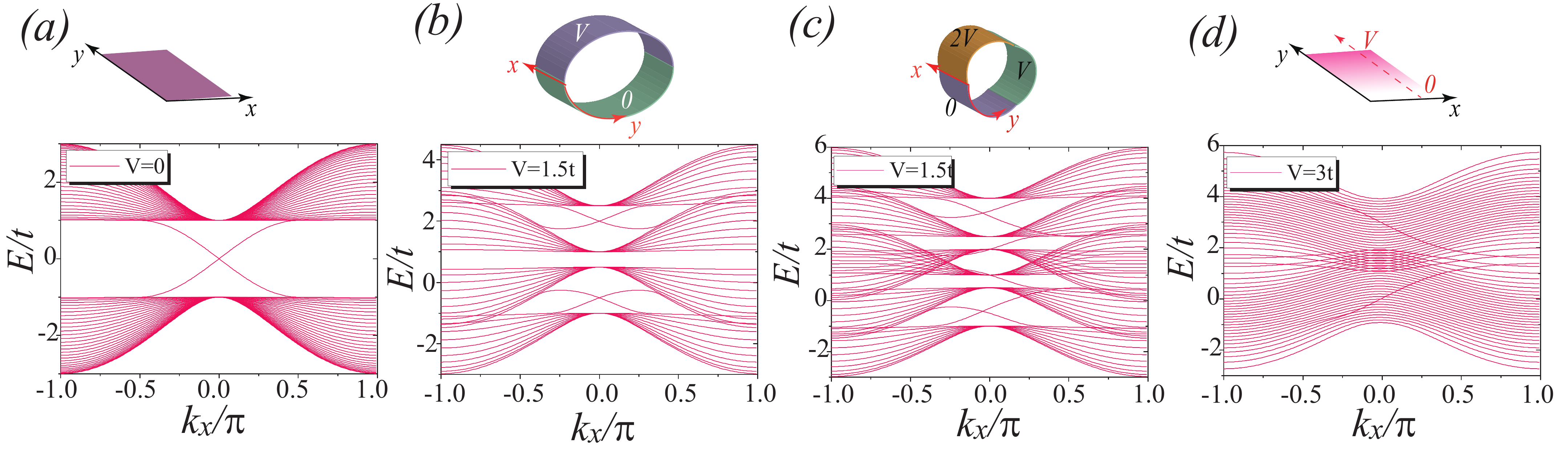}
\caption{The band structure of the QWZ lattice in four different cases.
(a) The band structure with voltage $V=0$ under open boundary condition.
(b) The band structure with voltage $V=1.5t$ under periodic boundary condition. The voltage along the $y$ direction is in the single step function form as $V_{\mathbf{i}} = V\Theta(i_y-N/2)$.
(c) The band structure with voltage $V=1.5t$ under periodic boundary condition. The voltage along the $y$ direction is in the double step function form as $V_{\mathbf{i}} = V\Theta(i_y-N/3) + V\Theta(i_y-2N/3)$.
(d) The band structure with voltage $V=3t$ under open boundary condition. The voltage distribution along the $y$ direction is in a linear function form as $V_{\mathbf{i}} = i_y V/N$ where $i_y$ is in the range of $[1, N]$.
Other parameters are the same as those in Fig. \ref{S0}.
}\label{S6}
\end{figure}

(1) Fig. \ref{S6}(a) shows the band structure with voltage $V=0$ under open boundary condition. The presence of the edge states inside the bulk gap shows the topological nature of the QWZ lattice.

(2) Fig. \ref{S6}(b) is the band structure with voltage $V=1.5t$ under periodic boundary condition. The voltage applied along the $y$ direction is $V_{\mathbf{i}} = V\Theta(i_y-N/2)$. It is shown that the band structure could be separated into two parts, and the energy difference between these two parts is about $V=1.5t$. Besides, the edge-state-like chiral interface states are also shown in the figure. Although the bulk gap decreases with the increasing of $V$, the bulk state is localized when Anderson disorder is presented. Thus, only the chiral interface state can still contribute to the conductance, which is consistent with our previous numerical results.

(3) Fig. \ref{S6}(c) is the band structure with voltage $V=1.5t$ under periodic boundary condition. Now the voltage applied along the $y$ direction is in the multi-step function form as
$V_{\mathbf{i}} = V\Theta(i_y-N/3) + V\Theta(i_y-2N/3)$. The band structure here is similar to that in Fig. \ref{S6}(b), except for the presence of a third part whose corresponding energy shift is $2V$. Due to the presence of the third part, the energy region for the topological interface states is enlarged, hence the multi-step electric potential contributes to the enhancement of the topological ``bulk gap''.

(4) The most important result is Fig. \ref{S6}(d), which shows the band structure with voltage $V=3t$ under open boundary condition. The voltage applied here is along the $y$ direction and in a linear function form as $V_{\mathbf{i}} = i_y V/N$. The linear bias makes the bulk band much more flat so that it benefits the Anderson localization.  The chiral interface state is still presented, although there is no global band gap.


\begin{figure}[b]
\includegraphics[width=8cm]{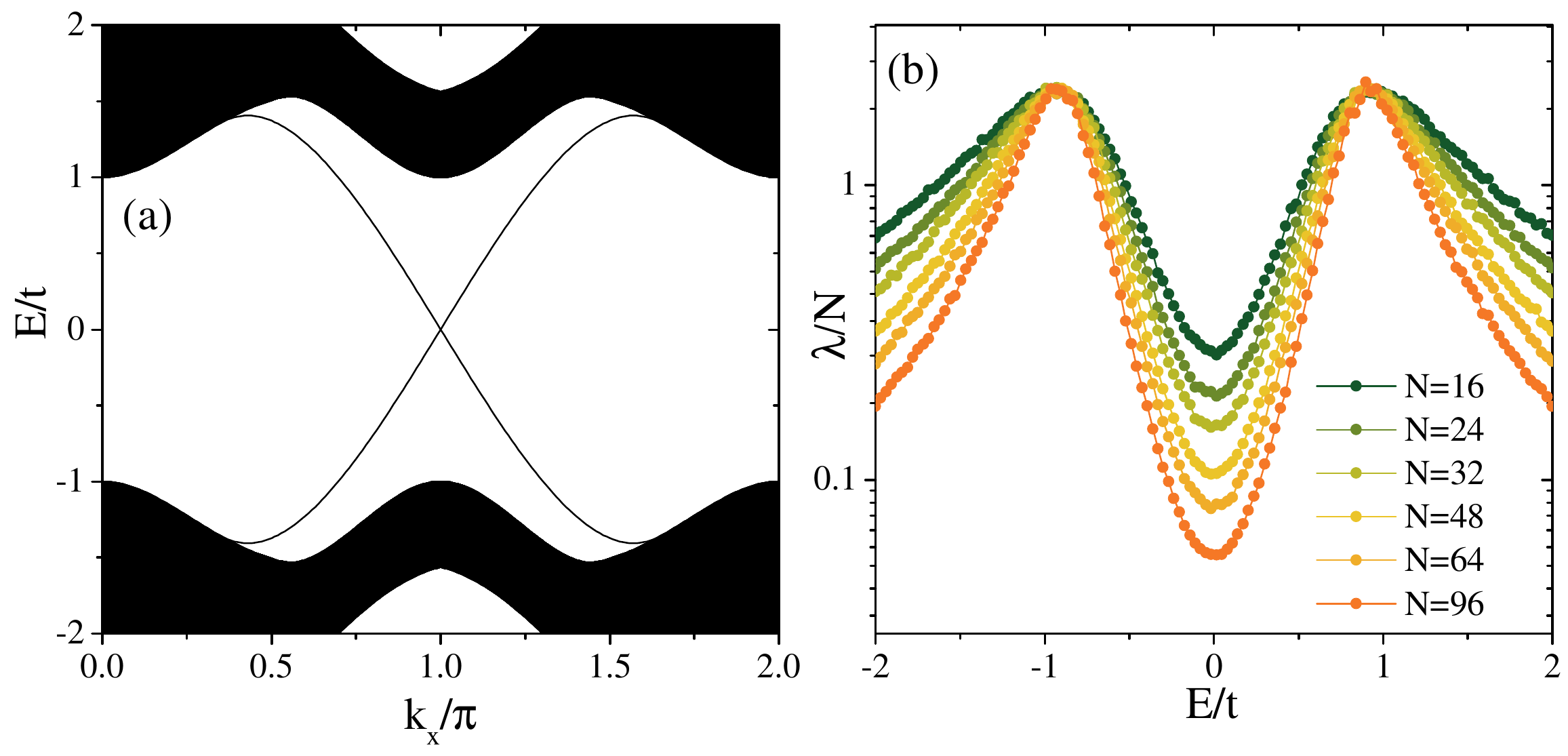}
\caption{(a) The band structure of the Haldane model.
(b) The corresponding renormalized localization length $\lambda/N$. All the parameters are the same as those in Fig. (a-1) of Ref. \cite{Sr1}.
}\label{S7}
\end{figure}


\section{S5. The Haldane model}

\begin{figure}[t]
\includegraphics[width=8cm]{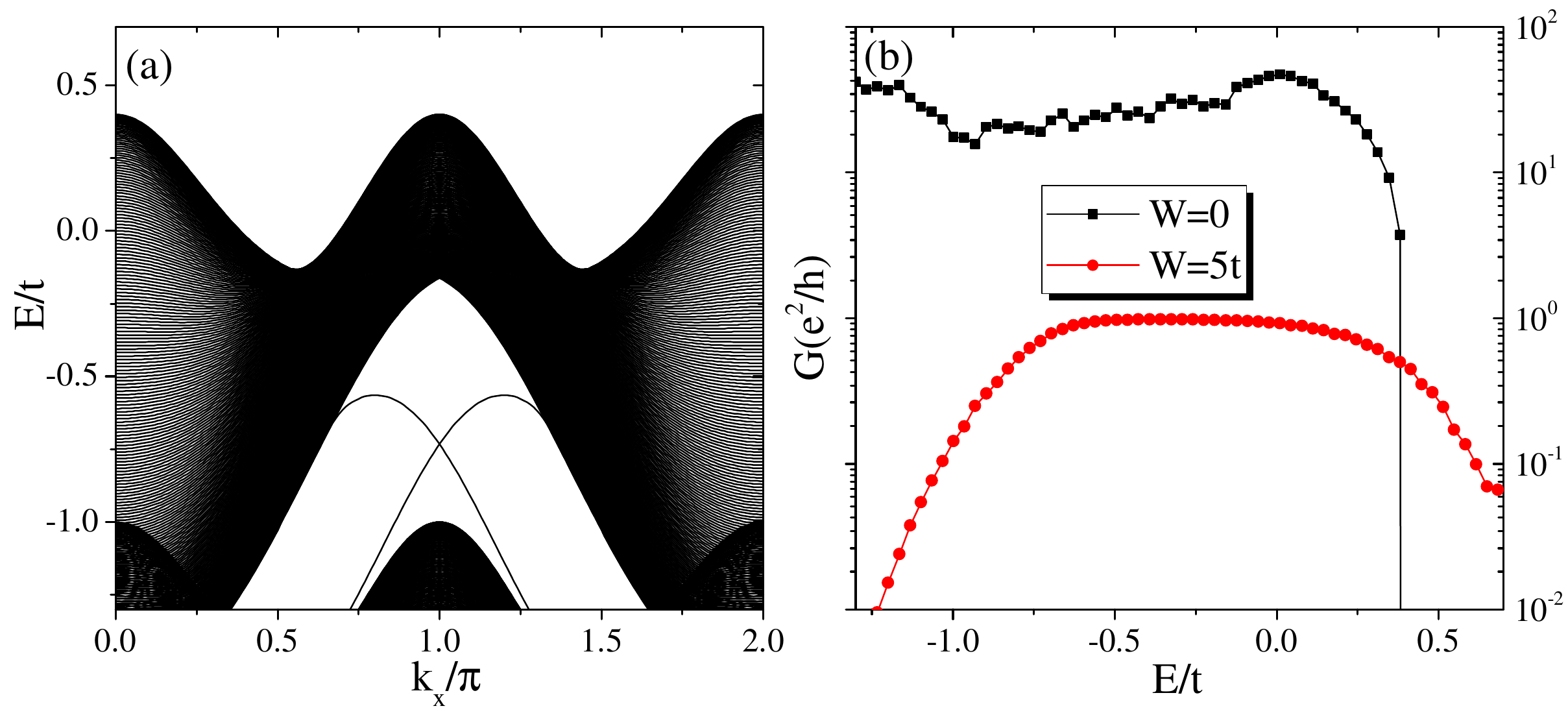}
\caption{(a) The band structure of the Haldane model under periodic boundary condition with applied voltage as $V_{\mathbf{i}} = V \Theta(i_y-N/2)$ where $V$ is fixed at $V=1.4t$.
(b) The corresponding conductance versus the Fermi energy $E$ for different disorder strength $W$. All the parameters are the same as those in Fig. (a-1) in Ref. \cite{Sr1}.
}\label{S8}
\end{figure}

In this section, we study the metal-insulator transition of the Haldane model \cite{sHaldane}. To demonstrate that the main results in this Letter is independent of the specific choice of the parameter sets, the parameters here for the Haldane model is chosen as the same as those in Ref. \cite{Sr1}. The band structure for the Haldane model is shown in Fig. \ref{S7}(a), which proves itself as a Chern insulator in the clean limit.
By using the transfer matrix method \cite{Scal2}, we also calculate the renormalized localization length $\lambda/N$ for cylindrical Haldane model lattices. In general, there could be three different behaviours for the renormalized localization length $\lambda/N$: (1) If $\lambda/N$ decreases with the increasing of $N$, it indicates an insulator behaviour; (2) For metallic phase, $\lambda/N$ increases with the increasing of $N$; (3) If $\lambda/N$ is invariant with the increasing of $N$, the transition points can be determined.
As shown in Fig. \ref{S7}(b), there is no metallic phase since $\lambda/N$ does not increase with the increasing of $N$. In addition, the transition point separating two different insulator phases (the normal insulator and the Chern insulator phases) is a point rather than a region. Since the localization length $\lambda$ for the phase transition points is of the same order as the sample size $N$, the conductance for these areas is difficult to reach the quantized value even in the thermodynamic limit $N\rightarrow \infty$.
In fact, as shown in Fig. 3(a) of the main text, a region with non-quantized conductance close to the phase transition point is presented between the two quantized plateaus. More details is going to be presented in the following section.

We replot the band structure of the cylindrical Haldane model lattice, which has been given in Ref. \cite{bulk}. The voltage applied here is in the form of $V_{\mathbf{i}} = V\Theta(i_y-N/2)$ with $V=1.4t$. As shown in Fig. \ref{S8}(a), the band structure is similar to the result shown in Fig. \ref{S6}(b), where the chiral interface states are mixed with the bulk states. The conductance for the clean sample is very large due to the contribution from the bulk states [Fig. \ref{S8}(b)]. In contrast, the conductance decreases when disorder is presented ($W=5t$), and a plateau with $G=e^2/h$ is obtained. These results once again show that our proposal for CIs is independent of the model as well as the Hamiltonian details.

\begin{figure}[b]
\includegraphics[width=15cm]{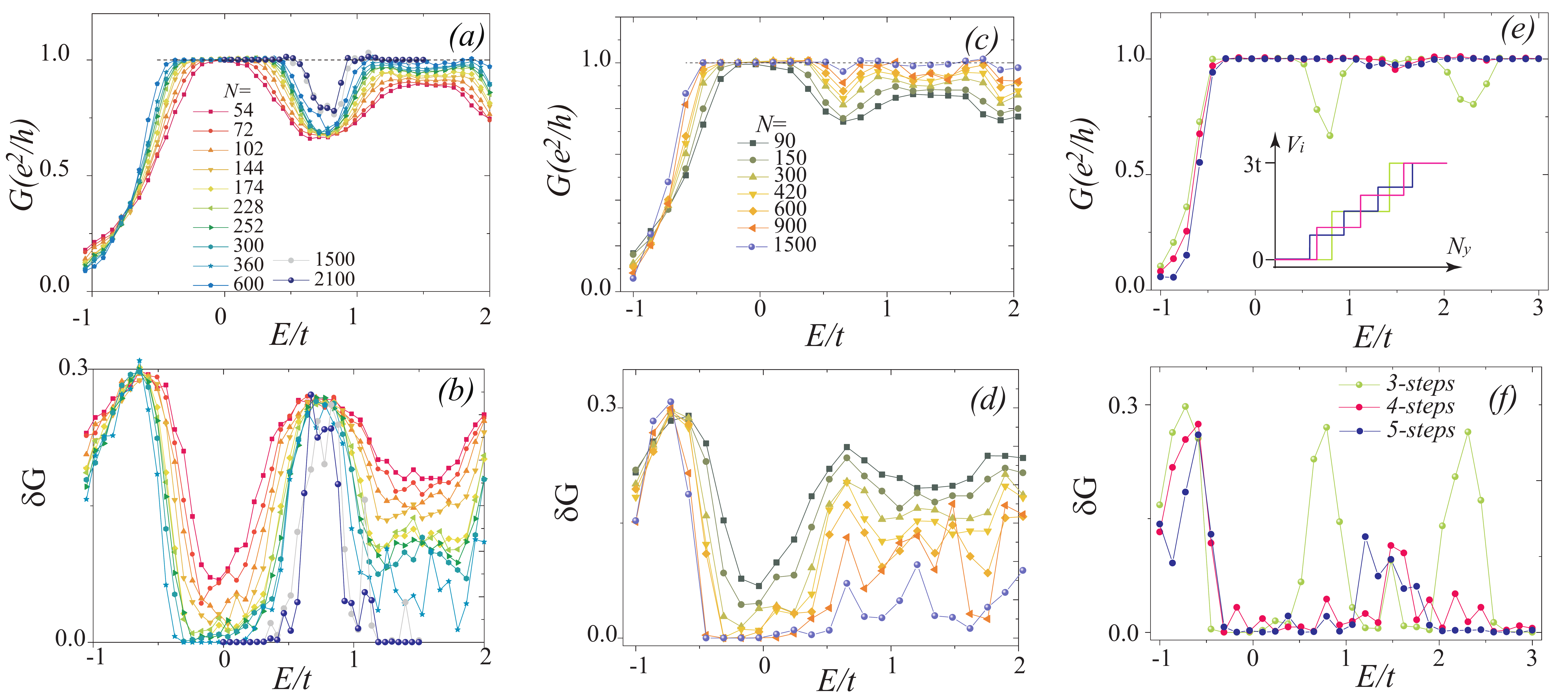}
\caption{(a), (b) The conductance $G$ and its corresponding conductance fluctuation $\delta G$ of Fig. 3(a) in the main text.
(c), (d) The conductance $G$ and its corresponding conductance fluctuation $\delta G$ for another samples with $V=1.3t$, while the other parameters are the same as those in (a).
(e), (f) The conductance $G$ and its conductance fluctuation $\delta G$ for different multi-step electric potential profile applied. The inset in (e) shows the corresponding electric potential profiles along the $y$ direction. The sample size is chosen as $N=900$ and the disorder strength is fixed at $W=3.5t$.
}\label{S9}
\end{figure}


\section{S6. The conductance fluctuation of Fig. 3(a) in the main text and the non-quantized conductance near the phase transition points}

The conductance $G$ and its corresponding fluctuation $\delta G$ of Fig. 3(a) in the main text are shown in Fig. \ref{S9}(a), (b), in which more conductance curves are also plotted. Moreover, the periodical boundary condition is adopted in the calculation for Fig. \ref{S9}(a), (b). To realize a perfect plateau near $E > -0.7t$, we have tried our best to enhance the sample size $N$. However, the dip between the two plateaus still exist. It is challenging to further increase the sample size $N$ since the time-cost for calculating the conductance curve with $N=2100$ is already about one and a half month. Thus, we pay more attention to the scaling of the conductance, which implies that there should be a broad $G=e^2/h$ plateau when $N\rightarrow \infty$.

We also notice that the conductance increases much slower when $E\approx 0.8t$. Based on the study of the metal-insulator transition in the Haldane model, the localization length $\lambda$ for the phase transition points is of the same order as the sample size $N$. Therefore, it is much more difficult to obtain the quantized conductance when the Fermi energy $E$ is close to the phase transition points. In other words, the Chern number is not quantized as $0$ or $1$ for such points even in the limit of $N\rightarrow \infty$. Thus, the chiral interface states are absent at these points even in the thermodynamic limit.
Furthermore, the sample size should be too large (beyond our computing power) when the Fermi energy is close to these transition points. Fortunately, the phase transition point is only a single point in the disordered CI so that it has little effect on the continuity of the quantized plateau when $N$ is large enough. As a consequence, our main conclusion for the chiral interface states and its related quantized transport still hold.

We also present the conductance and its corresponding fluctuation for CI samples with $V=1.3t$ [Fig. \ref{S9}(c), (d)]. It is obvious that the conductance plateau is also presented in such case. Besides, the conductance plateau here is much more smooth than the case of $V=1.5t$, which implies that the non-quantized conductance region near the phase transition points can be reduced by finding an appropriate parameter set.
Moreover, such non-quantized conductance region related to the phase transition points can be further eliminated by imposing an electric potential with multi-step form. For instance, as shown in Fig. \ref{S9}(e), a boarder and smoother plateau can be obtained by applying a multi-step electric potential along the $y$ direction [inset of Fig. \ref{S9}(e)], where the maximum and the minimum value of such potential is still fixed at $3t$ and $0$, respectively.
More importantly, by applying such multi-step potentials, it can be seen that the dip shown in Fig. \ref{S9}(a) is eliminated and the corresponding conductance fluctuation is also reduced at the same time.


\section{S7. The realization of the chiral interface states in LC circuits}

\begin{figure}[t]
\includegraphics[width=14cm]{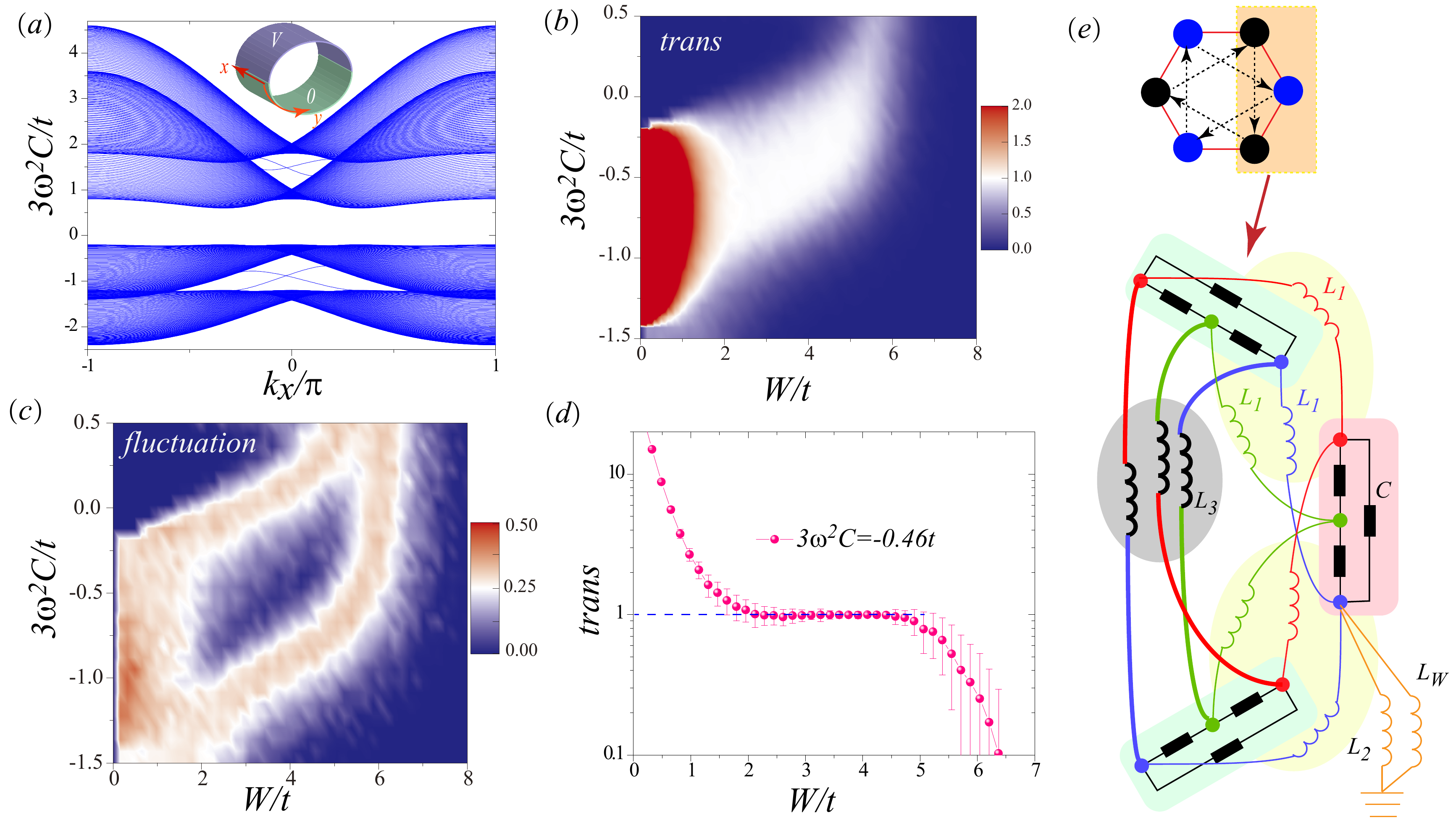}
\caption{(a) The dispersion of the Haldane model under periodical boundary condition which can be realized in the LC circuits. The nearest neighbour hopping strength is $-t$, the next nearest neighbour hopping strength is $t_2=-0.2t$, and the U(1) phase is fixed at $\phi=2\pi/3$. The electric potential applied here is the same as that in Fig. 2 in the main text with $V=t$.
(b), (c) The transmission coefficient (``trans'') and its corresponding fluctuation as functions of disorder strength $W$ and $3\omega^2C$, respectively. The sample size here is $N=100$.
(d) A typical plot of the transmission coefficient versus disorder strength $W$ with $3\omega^2C=-0.46t$. (e) The LC circuit structure for realizing the Haldane model.
}\label{S10}
\end{figure}

 In this section, we discuss the realization of our proposal in LC circuit. Following the previous study \cite{LC}, the nearest neighbour hopping strength of the Haldane model is fixed at $-t$, the next nearest neighbour hopping strength is set as $-0.2t$, and the U(1) hopping phase is chosen as $2\pi/3$. We first exhibit the band structure when a step potential with $V=t$ is presented [Fig. \ref{S10}(a)]. One can see that the dispersion of the LC circuit structure is similar to the previous results. Next, we calculate the transmission coefficient with respect to the disorder strength $W$ and $3\omega^2C$. A quantized transmission coefficient can be realized [Fig. \ref{S10}(b), (c)], where a typical ``trans''$\sim$$W$ plot is shown as Fig. \ref{S10}(d) with $3\omega^2C=-0.46t$. Since the fluctuation vanishes when the transmission coefficient is quantized at $1$, it is reasonable to claim that the topological interface states are also presented in such an LC circuit system.

Now we briefly present the circuit structure \cite{LC} realizing the honeycomb Haldane lattice. For simplicity, now we only pay attention to the orange rectangle [upper part of Fig. \ref{S10}(e)] in the honeycomb lattice, which contains three atoms, two nearest neighbour hopping term (solid red line), and one next nearest neighbour hopping term (dashed black line).  
Each of these three atoms corresponds to the circuit structure shown in a rectangle in the lower part of Fig. \ref{S10}(e), where each rectangle contains three circuit nodes. These three nodes are firstly connected by three capacitors with capacitance $C$ to form a closed ring.
The light green (gray) ellipse  denotes the nearest neighbour (next nearest neighbour) hopping term, which is realized by inductors with inductance $L_1$ ($L_3$). More importantly, the  $2\pi/3$ U(1) phase can be realized by connecting these inductors in a cross-connected fashion as shown in the lower part of Fig. \ref{S10}(e) \cite{LC}.
In addition, each node is grounded with two inductors with different inductances $L_2$ and $L_W$ [For simplicity, only the grounding inductors for one node is plotted in Fig. \ref{S10}(e)]. Specifically, the on-site energy (the voltage potential $V$) is inversely proportional to $L_2$ whose inductance is in a high accuracy. Meanwhile, the Anderson disorder can be achieved by $L_W$ whose inductance is in a sufficient error.
Theoretically, all the physical quantities are realizable in the circuit. One should also notice that the Green's function and the wavefunction corresponds to the impedances and the alternating current (AC) voltage signals in the circuits, respectively \cite{LC}. It suggests that the topological features of the interface states can be detected in such a circuit structure.

Experimentally, the parameters can be chosen as follows. The nearest neighbor and the next nearest neighbor hopping terms are related to the inductances as $L_1=\frac{1}{t}$ and $L_3=\frac{5}{t}$. Furthermore, the accurate inductance of the grounding inductors simulating the voltage potential satisfies $\frac{1}{V}=\frac{1}{t}$. The Anderson disorder with strength $W\approx 3.5t$ can be realized by using inductance $L_W\approx \frac{0.0577}{t}$ with an error ratio $y=0.1$. In addition, the total energy shift is approximately $ \frac{3}{L_1} +\frac{6}{L_3} + 17.5t$.
\begin{figure*}[b]
    \centering
	\includegraphics[width=0.8\textwidth]{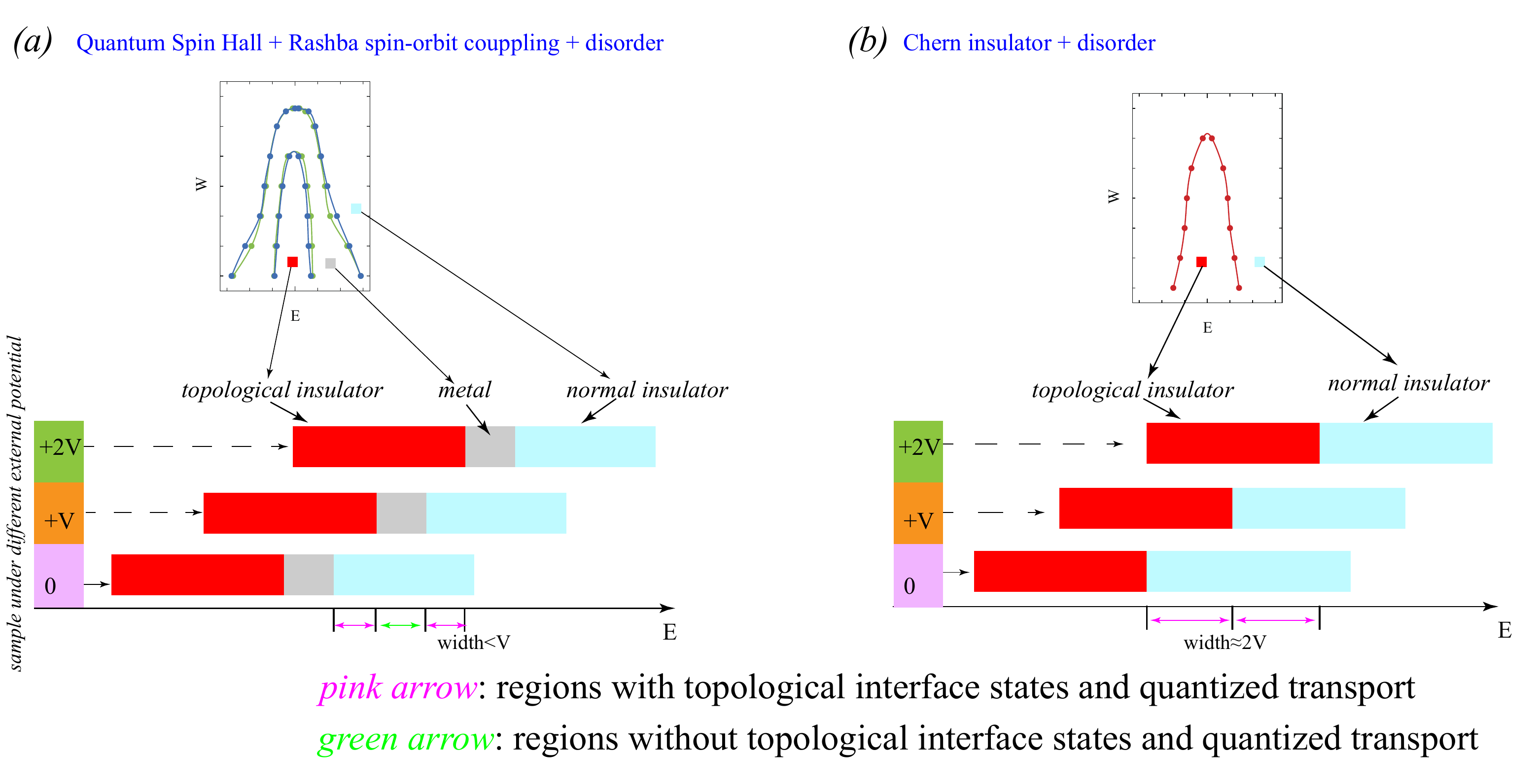}
\caption{(Color online).
(a) The upper panel: typical phase diagram with respect to the disorder strength $W$ for quantum spin Hall insulator (QSHI) with Rashba spin-orbit coupling. With the varying of the Fermi energy $E$,  the topological insulator phase (QSHI), metallic phase, and the normal insulator phase can be presented in such a disordered system.
The lower panel: a schematic diagram for different parts of the biased sample. The pink arrows represent the energy windows in which the interface states with quantized conductance is presented.
(b) is similar to (a) except for the QSHI being replaced by the Chern insulator. The phase diagrams are adapted from Ref. \cite{Sr1}.
}\label{metal}
\end{figure*}
Therefore, the Fermi energy is about $3\omega^2C\approx 21.24 t$, where $\omega$ is the frequency of the AC field and $C$ is the capacitance.
If the parameters are chosen as $L_1=1{\rm mH}$, $L_3=5{\rm mH}$ and $C=10{\rm \mu F}$, then the accurate grounding inductor is $L_2=1{\rm mH}$, and the disordered grounding inductors are in the inductance of ${L_W \approx 57~{\rm \mu H}}$ with the AC frequency $\omega\approx 26.6~{\rm KHz}$.

\section{S8. The influence of the metallic phase on our proposal}

 We clarify the influence of the metallic phase on our proposal.
Since disorder is inevitably presented in realistic samples, the metallic behavior should be described by the Anderson localization theory instead of the band theory.
The Anderson phase transition is universal and only depends on the system's spatial dimension and its symmetry ensemble according to the random matrix theory.

Disordered quantum spin Hall sample with Rashba spin-orbit coupling belongs to the symplectic ensemble. In such a system, as shown in Fig. \ref{metal}(a), topological insulator phase, metallic phase, as well as the normal insulator phase could be presented with the varying of the Fermi energy $E$.
The presence of the metallic phase will destroy the quantized conductance plateau in the corresponding energy region, which is marked by the green arrow in Fig. \ref{metal}(a).
Moreover, the width of the energy region with quantized conductance here [pink arrows in Fig. \ref{metal}(a)] is not determined by the external potential $V$.
Thus, such result is in stark contrast to the main consequence in our proposal, where the latter is based on the disordered Chern insulators belonging to the unitary class. Such symmetry ensemble only supports topological insulator phase (Chern insulator) and normal insulator phase [see Fig. \ref{metal}(b)], while the metallic phase is absent.
In short, our proposal is specific to the samples in the unitary class with two spatial dimensions [{\it e.g.} Chern insulators].

\section{S9. The evolution of chiral interface states versus Fermi energy}

\begin{figure*}[b]
    \centering
	\includegraphics[width=0.9\textwidth]{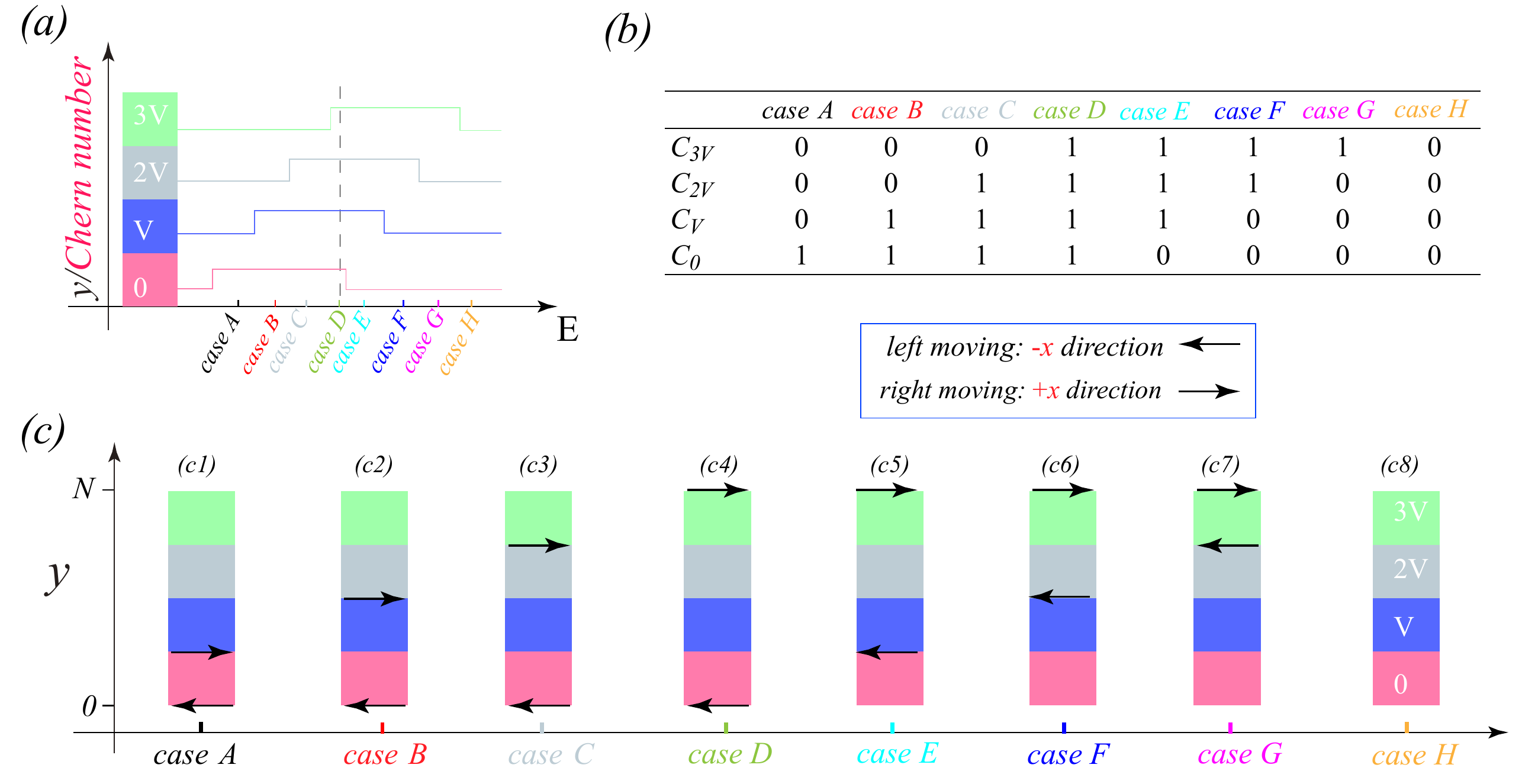}
\caption{(Color online).
(a) The Chern number for different parts of the sample in a four-step electric potential. The Chern number curves and the corresponding areas of the sample in (c) are marked by the same color. $Case$~$A$ to $case$~$H$ correspond to eight different Fermi energies.
(b) The set of Chern numbers $(C_{0},C_{V},C_{2V},C_{3V})$ for the areas with different electric potentials $(0,V,2V,3V)$ in the cases of different energies as shown in (a).
(c) The spatial positions of the interface channels (marked by black arrow) in the $y$-direction. The arrow's direction represents the right/left moving channel along the $+x$/$-x$ directions. $c1$ to $c8$ correspond to $case$~$A$ to $case$~$H$, respectively, as marked in the figure.
}\label{R2}
\end{figure*}

We take a sample in a four-step-potential as an example. The voltage distribution and the corresponding Chern number curves for different parts of the sample are plotted in Fig. \ref{R2}(a). For simplicity, we assume that $0 < V < \frac{\mathrm{mobility~gap}}{4}$. There could be eight different cases for different choice of the Fermi energy, which are marked by $case$~$A$ to $case$~$H$ with different colors. Fig. \ref{R2}(b) lists the set of Chern numbers $(C_{0},C_{V},C_{2V},C_{3V})$ for the sample region with different potentials $(0,V,2V,3V)$. According to the topological band theory, a chiral interface state emerges if the Chern number at the two sides of this interface differs by one. Hence, the chiral interface states for different cases can be obtained [as shown in Fig. \ref{R2}(c)] based on the Chern numbers shown in Fig. \ref{R2}(b).
It is obvious that $C_0=1$ remains unchanged from $case$~$A$ to $case$~$D$.
The nontrivial region becomes more extensive with the increase of the Fermi energy and finally extends to the entire sample in $case$~$D$. The backward chiral edge state always stays at the bottom of the sample, while the forward chiral interface state shifts along the $y$-direction step by step with the increase of the Fermi energy and finally reaches the top edge.
From $case$~$D$ to $case$~$G$, similarly, $C_{3V}=1$ remains unchanged while the area of the non-trivial region decreases. Finally in $case$~$H$ where the Fermi energy is large enough, the entire sample becomes topologically trivial so that the chiral interface state vanishes.

 To sum up, the disorder-induced localization in the CIs leads to the ``rectangular profile'' of the Chern number curves. At the same time, the sample is divided into parts with different potentials due to the external voltage. The combination of these two mechanisms results in the penetration of the chiral states into the bulk.

\section{S10. The influence of the potential slope}

\begin{table}[h]
\centering
\caption{ Sample sizes and the corresponding electron mobilities $\mu$ for different experimental samples. The references shown in this table are given in the main text.}\label{T1}
\begin{tabular}{l|c|c|c|r}
\hline
\hline
Material                           &  width($\mu \mathrm{m}$)  &  length($\mu\mathrm{m}$)     &  $\mu(\mathrm{cm}^{2}\mathrm{V}^{-1}\mathrm{s}^{-1})$   & Reference\\
\hline
Cr$_{0.15}$(Bi$_{0.1}$Sb$_{0.9}$)Te$_3$          &  50-200          &   50-400     & $\le 10^3$                    & Chang et al. Ref~[\b{4}]   \\
(Bi$_{0.29}$Sb$_{0.71}$)$_{1.89}$V$_{0.11}$Te$_3$  &  $500$    &    $2000$           & $130$                     & Chang et al. Ref~[\b{5}]   \\
Cr$_{x}$(Bi$_{1-y}$Sb$_y$)$_{2-x}$Te$_3$         &  200          &   600     & $270$                    & Checkelsky et al. Ref~[\b{7}]   \\
MnBi$_2$Te$_4$         &  20          &   $\le 40$     & 100-700                    & Deng et al. Ref~[\b{8}]   \\
MnBi$_2$Te$_4$         &  20          &   $20$     & $74$                    & Liu etal. Ref~[\b{9}]   \\
\hline
\hline
\end{tabular}
\end{table}

The plateau will be destroyed if the slope is too large. However, the detailed inspection below demonstrates that such constraint could barely restrict the increase of the desired energy window in realistic conditions.

The distance between two counterpropagating chiral edge states is not only determined by the slope of the voltage. As shown in Fig. \ref{R2}, for the multi-step potential case, the distance between two counterpropagating chiral edge states is always an integer multiple of the width of the potential steps.
For linear biased samples, the evolution of the counterpropagating chiral edge states is also similar to the case shown in Fig. \ref{R2}. For simplicity, we suppose the localization length (along the $y$-direction) of the topological edge (interface) states to be $\lambda$. To ensure that the quantized conductance plateau will not be interrupted, the critical electric field (voltage slope) $E_c$ in such a sample should satisfy the relation as:
\begin{equation}
2\lambda E_c\approx \Delta,
\end{equation}
where $\Delta$ is the effective energy gap of the sample. Hence, for a sample whose width along the $y$-direction is $L_y$, the maximum external potential allowed is about $V_m=E_cL_y \approx \frac{ L_y\Delta}{2\lambda}$. Thus, correspondingly, the effective energy gap can be enlarged by about $\frac{L_y}{2\lambda}$ times.
In a typical quantum anomalous Hall system, the localization length is $\lambda\approx\frac{\hbar v_F}{\Delta}\sim10$-$100\mathrm{nm}$ \cite{Axion}  where $v_F$ is the Fermi velocity.
The typical experimental sample widths $L_y$ (summarized in TABLE \ref{T1}) are much larger than $\lambda$. Thus, even a small electric field $E_l$ satisfying $E_l \ll E_c$ will significantly enlarge the quantized ``transport gap''.
Consequently, though the plateau could be destroyed when $V$ is large enough, the ``transport gap'' has already been significantly enhanced before the plateau is destroyed.

\section{S11. The influence of the smoothed potential}

\begin{figure*}[t]
    \centering
	\includegraphics[width=0.6\textwidth]{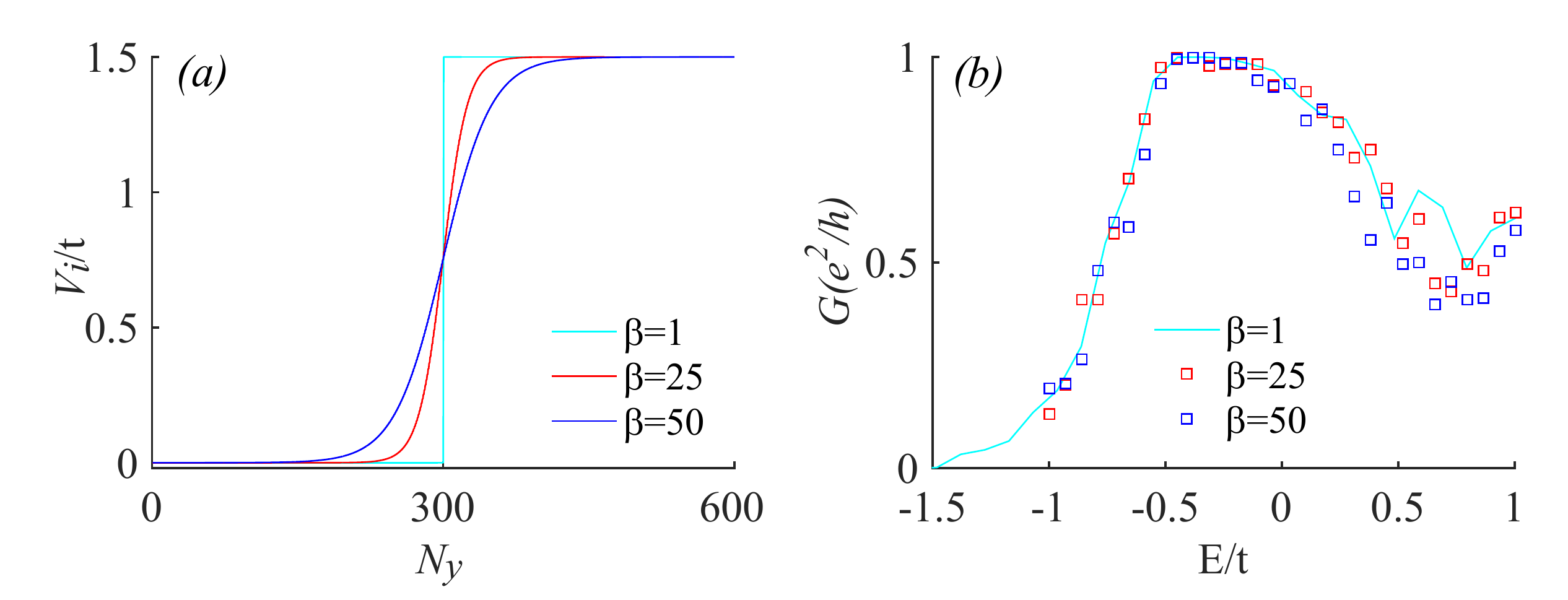}
\caption{(Color online).
(a) $V_i$ versus $N_y$ for different smoothness $\beta$ with $N=600$ and $V=1.5t$.
(b) The corresponding two-terminal conductance $G$ versus the Fermi energy $E$. Other parameters are the same as those in Fig. 2(a) in the main text.
}\label{R3}
\end{figure*}

 In the following, we replace the step potential with the following smoothed potential\cite{Sp}:
\begin{equation}
V_{i} = \mathrm{tanh}[(N_y-N/2)\beta]V/2+V/2,
\end{equation}
\noindent where $i\equiv N_y\in[0,N]$, $V$ is the bias, $\mathrm{tanh}(x)$ is the hyperbolic tangent function, and $\beta$ describes the ``smoothness'' of the potential profile. In our calculation, we fix $N=600$ and $V=1.5t$. As shown in Fig. \ref{R3}(a), the $V_i$-$N_y$ curve becomes more smooth with the increase of $\beta$. Fig. \ref{R3}(b) shows the corresponding conductance curve (with respect to the Fermi energy $E$) along the $x$-direction. It can be seen that such conductance curve is independent of the smoothness of the potential distribution $\beta$ along the $y$-direction.

\section{S12. The possibility of the realization of our proposal in $MnBi_2Te_4$}

\begin{figure*}[t]
    \centering
	\includegraphics[width=0.4\textwidth]{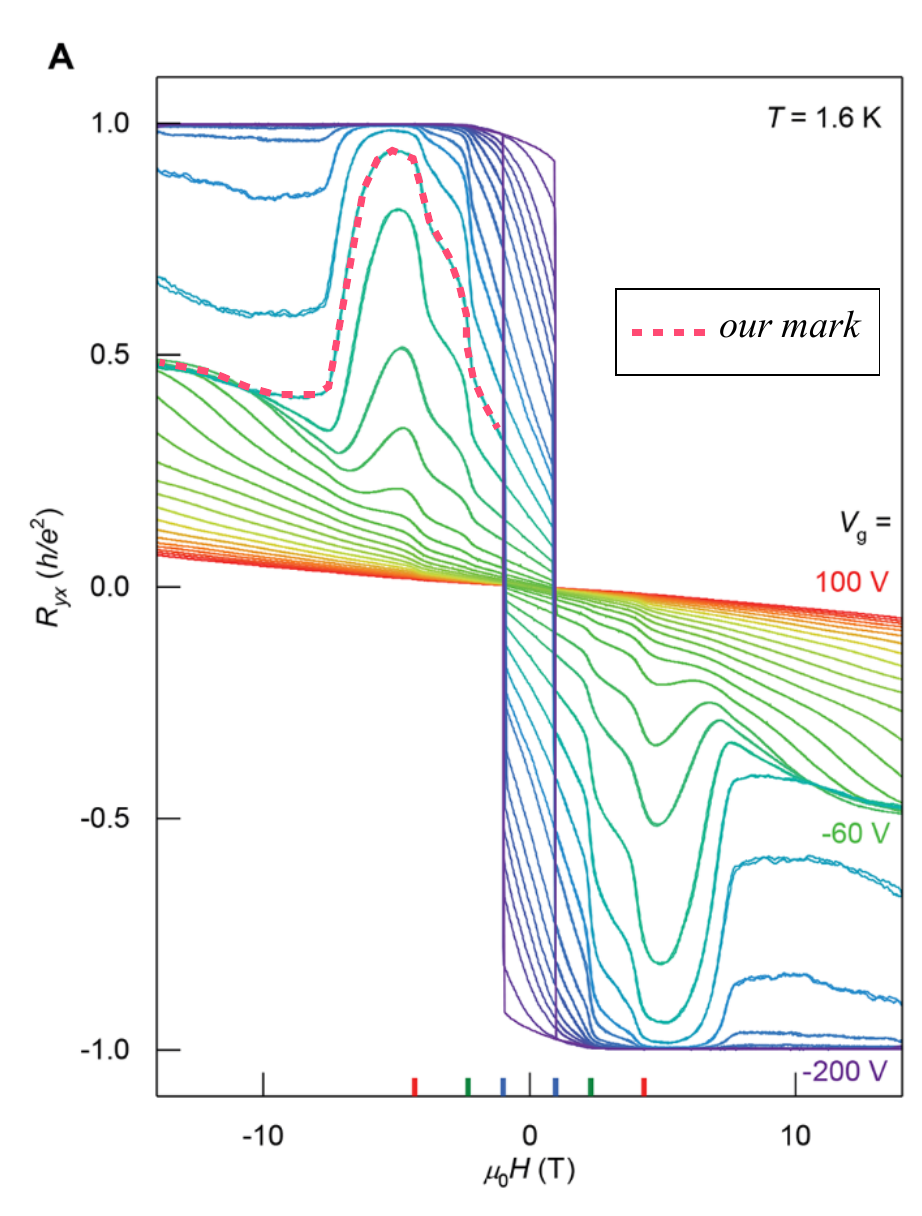}
\caption{(Color online).
The Hall conductance versus magnetic field $\mu_0H$ at different gate voltages $V_g$. This picture is adapted from experiment \cite{paper2}. The red dashed line is marked to highlight the data we discussed.
}\label{R0}
\end{figure*}

Significantly, a recent paper published in $Science$ \cite{paper2} has strongly suggested the existence of Anderson localization in MnBi$_2$Te$_4$ 
as well. In that paper, the authors obtained a quantized plateau with $R_{xy}\approx h/2e^2$ under strong magnetic field.
When the magnetic field decreases, a plateau with $R_{xy}\approx h/e^2$ appears (see the red dashed line in Fig. \ref{R0}).
In the case of strong magnetic field, the authors observed the coexistence of the quantum Hall (QH) effect with $R_{xy}\approx h/e^2$ and the Quantum anomalous Hall (QAH) effect with $R_{xy}\approx h/e^2$ in the same sample. It indicates that the Fermi energy is located at the bulk band since the bulk electrons form Landau levels under strong magnetic field.
The QH effect fades away with the decrease of the magnetic field, while the Fermi energy should still lies within the bulk band. At this time, the $R_{xy}$ from the QAH effect should not be quantized if the bulk band of the system is in the metallic phase. However, the $R_{xy}\approx h/e^2$ exhibited in Fig. \ref{R0} strongly suggests that the QAH still can be observed.
The only explanation is that the corresponding bulk states are localized in such a system. Consequently, such recent experimental observation implies that our proposal can be implemented in realistic materials.

\section{S13. The localization of inter face states}

\begin{figure*}[t]
    \centering
	\includegraphics[width=0.7\textwidth]{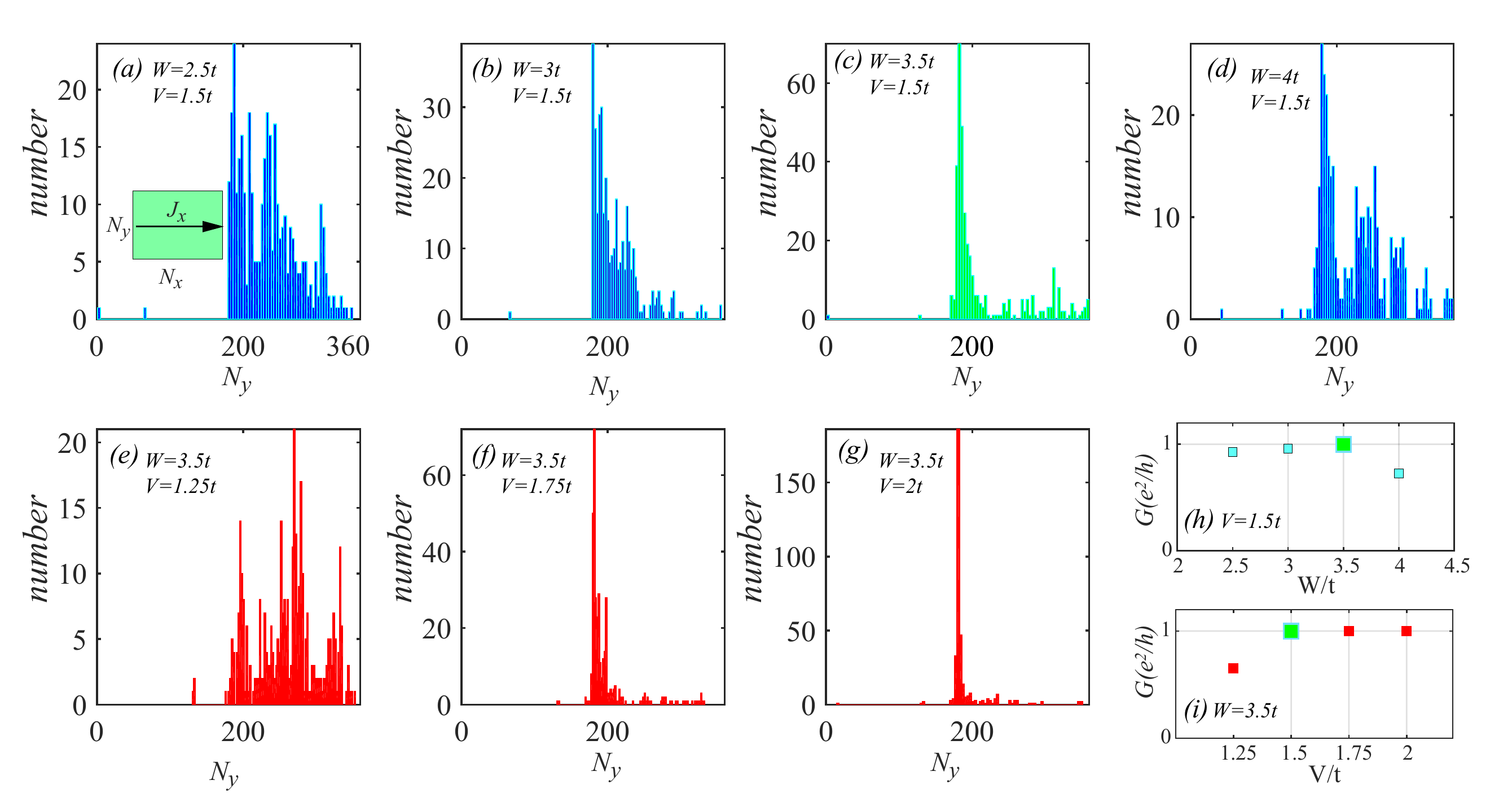}
\caption{(Color online).
(a)-(g) We pick $N_y$ of the maximum local-current density along $x$ direction ($J_x(N_x,N_y)$) for each slice $N_x\in[1,N]$. Then, the statistical histograms of $N_y$ for samples under different parameters are presented. The sample size is $N=360$, and the step potential along $y$ direction is $U_i=V\Theta(N_y-N/2)$ with $N_y\in[1,N]$. The periodical boundary along $y$ direction is adopted. The inset of (a) is a schematic diagram of the setup. (h) and (i) are the two-terminal conductance $G$ for (a)-(d) and (e)-(g), respectively. Other parameters have been given in the figures. The Fermi energy is set as $E_F=-0.1t$.
}\label{R23}
\end{figure*}

Because the localized bulk states has little effects on the local-current density along $x$ direction [$J_x(N_x,N_y)$ with $N_x, N_y\in [1,N]$, the current is also along $x$ direction], we use $J_x(N_x,N_y)$ to uncover the localization properties of the ``edge states" instead. Taking samples with one step potential as an example, we pick $N_y$ of the maximum $J_x(N_x,N_y)$ for each slice $N_x\in[1,N]$. Then, the statistical histograms of $N_y$ for samples under different parameters are presented in Fig.~\ref{R23}. The corresponding conductance $G$ is also given. It clearly shows that the disorder strength $W$ and the potential $V$ will influence the topological interface states' localization.
In short, if the conductance $G$ is closer to the quantized value $e^2/h$, the topological channel is more localized.

\end{widetext}


\begin{thebibliography}{99}
\bibitem{footnote1} For a cylinder sample, there are two interfaces.  The Chern number difference $\pm 1$ along the $y$-direction leads to interface states with opposite chiralities. When a positive bias is applied, only right-moving chiral interface state (the corresponding Chern number difference is $1$) contributes to the quantized conductance.
\end{thebibliography}
\end{document}